\documentclass[11pt]{article}
\usepackage{amsfonts}
\usepackage{amsmath}

\begin{document}

\newcommand{\Bergman}{{\widetilde{\Cset \Prm^2}}}
\newcommand{\AdS}{\mathrm{AdS}}
\newcommand{\HH}{\Hset\mathrm{H}^1}
\newcommand{\vone}{\text{v}_1}
\newcommand{\vtwo}{\text{v}_2}
\newcommand{\Ub}{\bar{U}}
\newcommand{\Rmn}{{\mathcal R}_{\mu\nu}}

\newcommand{\half}{\frac{1}{2}}
\newcommand{\diff}{\mathrm{d}}
\newcommand{\p}{\partial}
\newcommand{\nceq}{\doteq}
\newcommand{\ra}{\rightarrow}
\newcommand{\lra}{\leftrightarrow}
\newcommand{\comment}[1]{}

\newcommand{\Rcal}{{\mathcal R}}
\newcommand{\Acal}{{\mathcal A}}
\newcommand{\Dcal}{{\mathcal D}}
\newcommand{\Mcal}{{\mathcal M}}
\newcommand{\Ncal}{{\mathcal N}}
\newcommand{\Lcal}{{\mathcal L}}
\newcommand{\Scal}{{\mathcal S}}

\newcommand{\Zset}{{\mathbb Z}}
\newcommand{\Rset}{\mathrm{I}\kern-.18em\mathrm{R}}
\newcommand{\Cset}{{\,\,{{{^{_{\pmb{\mid}}}}\kern-.47em{\mathrm C}}}}}
\newcommand{\Hset}{\mathrm{I}\kern-.18em\mathrm{H}}

\newcommand{\Prm}{{\mathrm P}}
\newcommand{\Erm}{{\mathrm E}}
\newcommand{\Frm}{{\mathrm F}}
\newcommand{\Grm}{{\mathrm G}}
\newcommand{\Orm}{{\mathrm O}}
\newcommand{\Urm}{{\mathrm U}}
\newcommand{\Srm}{{\mathrm S}}
\newcommand{\SO}{\mathrm{SO}}
\newcommand{\Sprm}{\mathrm{Sp}}
\newcommand{\SU}{\mathrm{SU}}

\newcommand{\gra}{\alpha}
\newcommand{\vt}{\vartheta}
\newcommand{\vf}{\varphi}
\newcommand{\ff}{\varphi}
\newcommand{\grl}{\lambda}
\newcommand{\grL}{\Lambda}

\newcommand{\sone}{\sigma_1}
\newcommand{\stwo}{\sigma_2}
\newcommand{\sthree}{\sigma_3}
\newcommand{\zone}{{z_1}}
\newcommand{\zoneb}{{\bar{z}_1}}
\newcommand{\ztwo}{{z_2}}
\newcommand{\ztwob}{{\bar{z}_2}}
\newcommand{\zp}{\zeta_{+}}
\newcommand{\zm}{\zeta_{-}}
\newcommand{\xb}{x}
\newcommand{\tr}{\theta_r}
\newcommand{\ts}{\theta_s}
\newcommand{\fr}{\varphi_r}
\newcommand{\fs}{\varphi_s}
\newcommand{\mn}{{\mu\nu}}

\numberwithin{equation}{section}

\begin{titlepage}

\begin{flushright}
 {\tt YITP-SB-02-50}
\end{flushright}

\mbox{}
\bigskip
\bigskip
\bigskip

\begin{center}
{\Large \bf Holography and Quaternionic Taub--NUT}\\
\end{center}

\bigskip  
\bigskip  
\bigskip  
\bigskip 
  
\centerline{\bf Konstantinos Zoubos}
  
\bigskip  
\bigskip  
\bigskip  
  
\centerline{\it C. N. Yang Institute for Theoretical Physics}
\centerline{\it State University of New York at Stony Brook}
\centerline{\it Stony Brook, New York 11794-3840}
\centerline{\it U. S. A.}
\bigskip

\centerline{\small \tt  
  kzoubos@insti.physics.sunysb.edu}  
  
\bigskip  
\bigskip  
\bigskip

\begin{abstract}  
  \vskip 4pt As a concrete application of the holographic 
correspondence to manifolds which are only asymptotically  Anti--de Sitter, we 
take a closer look at the quaternionic Taub--NUT 
space. This is a four dimensional, non--compact, inhomogeneous, riemannian manifold
 with the interesting property of smoothly interpolating 
between two symmetric spaces, AdS$_4$ itself and the coset $\SU(2,1)/\Urm(2)$.
Even more interesting is the fact that the scalar curvature of the induced 
conformal structure at 
the boundary (corresponding to a squashed three--sphere) changes sign as we 
interpolate between these two limiting cases.
Using twistor methods, we construct the bulk--to--bulk and bulk--to--boundary
propagators for conformally coupled scalars on quaternionic Taub--NUT.  
This may eventually enable us to calculate
correlation functions in the dual strongly coupled CFT on a squashed $\Srm^3$ 
 using the standard AdS/CFT prescription. 
\end{abstract}

\end{titlepage}

\setcounter{footnote}{0}  
\noindent

\section{Introduction}

 Since the formulation of the AdS/CFT conjecture in 
\cite{Maldacena98,Witten98,Gubseretal98}
there has been considerable effort in trying to check whether its validity extends to
situations more general than the original $\AdS_{d+1}\times\Srm^{D-d-1}$ cases (where $D$ is 10 
or 11). Assuming one chooses to remain in the limit where classical supergravity is applicable to 
some extent,
there are two obvious generalisations. The first one is to use a different
compact internal manifold (or orbifold). Since in this case we preserve the AdS part of the
spacetime the dual field theory remains conformal, but substituting internal
manifolds of different holonomies leads to less than maximal supersymmetry.
Examples of this type are rather numerous and we refer to \cite{Aharonyetal00}
for an exhaustive list of references. 

The other route is to replace the AdS factor of the compactification by a different 
non--compact Einstein manifold $X_{d+1}$. If $X_{d+1}$ asymptotically gives $\AdS_{d+1}$ as the 
boundary is approached, then the dual field theory is expected to be 
conformal only in the UV. (Examples of this are the AdS--Schwarzschild black
holes of \cite{Witten98,Witten98a}.) 
Other examples of non--AdS space--time manifolds include (in no particular order)
the domain--wall and domain--wall black hole solutions of 
\cite{Boonstraetal98} and \cite{CaiOhta00} respectively, pp--waves in AdS
\cite{Cveticetal99a,Brecheretal01}, 
the coset spaces considered in \cite{Britto--Pacumioetal99}, the de Sitter/CFT
correspondence (e.g. \cite{Strominger01}),  and cases 
where AdS is modded out by discrete groups, as in \cite{HorowitzMarolf98,Gao99}.
Finally, \cite{Taylor--Robinson00}
considers the implications for holography when one has manifolds whose boundary conformal 
structure is degenerate. 
 
 There also exists a method, called holographic renormalisation 
\cite{HenningsonSkenderis98,deHaroetal01} (see \cite{Skenderis02} for a review), that 
enables the systematic computation of renormalised
correlation functions for operators in the field theory dual to any asymptotically $\AdS$ 
space. (An Einstein manifold with negative cosmological constant 
is called ``asymptotically $\AdS$'' if, like $\AdS$ itself, the metric has a 
double pole as the boundary is approached, 
so that one obtains not a boundary metric, but a conformal class of metrics.
 See \cite{Skenderis01} for the precise definition.) This method has been applied to 
asymptotically $\AdS$ domain wall solutions of supergravity that describe  RG flows
in the dual field theory \cite{Bianchietal01,Bianchietal02}. Holographic renormalisation depends
 on a near--boundary analysis, in which the bulk fields are expanded in the radial
coordinate transverse to the boundary, and the field equations are solved term--by--term
in this expansion. However, this is a local analysis which cannot specify all the coefficients
in the series, in particular those that are related to non--local quantities like 
correlation functions. To do that one would like to have an \emph{exact} solution of the bulk field
equations, extending smoothly to the deep interior of the bulk manifold.

 Here we are concerned with the quaternionic Taub--NUT manifold (QTN from 
now on) which is a four--dimensional riemannian Einstein manifold, a limiting case of 
which corresponds to $\AdS_4$ (or, since we work exclusively with 
euclidean signature, quaternionic hyperbolic space $\HH$). The Weyl tensor of QTN is self--dual
(rather than zero as is the case for hyperbolic space) and so, unlike $\HH$, it is 
not conformally flat. However, like $\HH$, its metric has a double pole as one approaches the 
boundary, so it is asymptotically $\AdS$. We will consider the Dirichlet problem for a 
 scalar propagating in the fixed QTN background and compute exact bulk--to--bulk and 
bulk--to--boundary propagators in the conformally coupled case. This may eventually allow
us to calculate two-- and higher--point functions in the dual theory by applying the 
methods of holographic renormalisation.

 We should note that the QTN manifold has already been used in generalising the 
AdS/CFT correspondence in 
\cite{Hawkingetal99,Chamblinetal99,Emparanetal99,Mann99}, but all this previous work
has been concerned with calculating Casimir energies for the dual field 
theory. We believe that our work will take the correspondence one step further.

As is well known, the boundary of $\HH$ is the bi-invariant (``round'')
metric on $\Srm^3$:
\begin{equation} \label{roundsphere}
\diff s^2=\sone^2+\stwo^2+\sthree^2\;\;.
\end{equation}
This is of course a particular choice of representative for the conformal structure of
the boundary.  Since this metric is conformally flat, by a different choice of 
representative one usually writes the boundary as three--dimensional flat space $\Rset^3$
and, via AdS/CFT, proceeds to define a dual conformal field theory thought of as living
on this flat boundary.
 
In \cite{Pedersen86}, Pedersen applied earlier ideas of LeBrun \cite{LeBrun82} to
solve the problem of obtaining an Einstein
4--metric that has as conformal infinity the Berger sphere, i.e. the
$\SU(2)\times\Urm(1)$--invariant ``squashed'' three-sphere metric\footnote{Pedersen's
work can be thought of as an example of the subsequent theorem of Graham and Lee 
\cite{GrahamLee91} that the conformal class of a given metric on $\Srm^d$, 
sufficiently close to the round one,
can arise as the conformal infinity of a suitable $d+1$--dimensional metric on the ball
$B^{d+1}$. (The general problem of whether such metrics exist was posed in 
\cite{FeffermanGraham85}.)}:
\begin{equation} \label{Bergersphere}
\diff s^2=\sone^2+\stwo^2+\grl \sthree^2\;\;.
\end{equation}
The resulting complete Einstein manifold with negative cosmological constant 
turns out to be quaternionic Taub--NUT. Since the metric on QTN smoothly extends across
the boundary to the conformal structure represented by (\ref{Bergersphere}),
it is natural to expect that one can apply the AdS/CFT prescription to this case. 
We show that one can find (scalar) bulk--to--bulk and bulk--to--boundary propagators 
that  generalise the ones known for $\HH$ and thus in principle obtain 
correlation functions for certain operators in the dual (3+0) conformal
field theory living on the squashed three--sphere. However, since the squashed 
three--sphere is not conformally flat, we will be forced to consider the
dual CFT as living on a curved background. (Examples of AdS/CFT with curved
boundaries can be found in e.g. 
\cite{LandsteinerLopez99,Brecheretal01,KoyamaSoda01,CardosoLust02}.)

We do not expect that we might learn something about realistic
field theories in this fashion. For a start, we discuss only riemannian
metrics and it is not clear whether one can translate these results  
to the lorentzian regime. Also, we make no specific reference to 
any M-theory or string compactifications, although  one could
directly replace $\HH$ by QTN in the standard Freund--Rubin ansatz to
obtain e.g. a QTN$\times \Srm^7$ vacuum solution of 11--dimensional 
(euclidean) supergravity. We will also not touch the issue of stability, at 
least from the gravity side of the correspondence (but see below).

The point of view we take is that we might learn something about holography 
by examining a more general setting that is still tractable (because of the 
quaternionic geometry). In particular we note the following interesting 
features of holography on QTN that provided motivation for this work and, we believe,
merit further investigation:

\begin{enumerate}
\item Obviously the round (conformally flat) metric (\ref{roundsphere}) 
on $\Srm^3$ ($\grl=1$) has positive scalar curvature. However as we squash the 
three--sphere by increasing the parameter $\grl$, we reach a point ($\grl=4$)
where the scalar curvature becomes zero, and for $\grl>4$ it becomes negative
(this is  discussed e.g. in \cite{Duffetal86}, p. 38). 
It is well documented in the AdS/CFT literature (e.g. in 
\cite{SeibergWitten99,WittenYau99}\footnote{See also \cite{McInnes01}.}) 
that for the field theory 
dual to any negative curvature Einstein Manifold to be stable, it is
necessary that the induced boundary manifold have positive scalar curvature. 
So it is nice to be able to explore a case where we can smoothly move
from a positive to a negative scalar curvature boundary metric, and thus try
to understand the apparent instability of the boundary theory in bulk terms.
(Presumably the instability will have something to do with emission of a 
``large'' brane as in \cite{SeibergWitten99}.)
The point $\grl=4$ where the  boundary  has zero scalar curvature 
is also quite intriguing in its own right.

\item If we take our parameter $\grl$ to its limiting value of 
infinity, the Berger sphere degenerates and 
we end up with an effectively two dimensional theory. In 
$\cite{Britto--Pacumioetal99}$ it is claimed that the resulting CFT is
in fact well--behaved (being dual to a stable M--theory compactification
on $\SU(2,1)/\Urm(2)\times S^7$). 
 It is thus interesting to analyse how the transition from a 3d to a 
2d CFT is made, and how the apparent stability can be reconciled with
the point made above, since we are in the negative boundary scalar curvature 
regime (in fact, the scalar curvature is negatively infinite)\footnote{This issue has been raised
in \cite{Taylor--Robinson00}.}. 
\end{enumerate}

Our approach to calculating propagators is inspired by  
Page's construction of Green's functions for 
massless scalars propagating on gravitational multi--instantons \cite{Page79}. 
This work is essentially a generalisation of those results to 
the non--Ricci--flat case. In the next section we give an overview of the
quaternionic Taub--NUT manifold. In section 3 we specialise to the metric
we will use for QTN, which is the one found by Pedersen, and exhibit 
several interesting limiting cases. In section 4 we review known results
for Green's functions for these cases. The construction of the conformally 
coupled Green's function
for the Pedersen metric follows in sections 5,6,7. In section 8 we briefly discuss 
the boundary theory and we conclude with various open issues in section 9.

\section{Some facts about Quaternionic Taub--NUT}

In $4n$ real dimensions, we use the term quaternionic Taub--NUT (QTN$_n$) for the 
noncompact inhomogeneous quaternionic K\"ahler manifold that has 
euclidean Taub--NUT as its hyperk\"ahler limit. A quaternionic K\"ahler 
manifold is defined (e.g. \cite{Ishihara74,Salamon82}) to be a $4n$--dimensional 
oriented riemannian manifold with holonomy contained in $\Sprm(n)\cdot\Sprm(1)$ (which means
$(\Sprm(n)\times\Sprm(1))/\Zset_2$).
In four dimensions (which is all we consider here) the holonomy 
requirement $\Sprm(1)\cdot\Sprm(1)\cong \SO(4)$ is vacuous, so one defines 
quaternionic K\"ahler manifolds as those Einstein manifolds whose Weyl tensor
is self--dual (or anti--self--dual). Since we consider only the four dimensional
case, we write just QTN for QTN$_1$. 
 
Quaternionic K\"ahler manifolds have (positive or negative) cosmological
constant $\grL$, and reduce to Ricci--flat hyperk\"ahler manifolds in the
limit where $\grL$ is taken to zero. We are interested in noncompact manifolds,
which (by Myers' theorem) means  $\grL$ has to be non--positive.  
In four dimensions there are only two noncompact \emph{homogeneous} 
quaternionic K\"ahler spaces (which, actually, are also symmetric): 
Hyperbolic space and the noncompact version of 
$\Cset\Prm^2$, 
which we denote by $\Bergman$. Being homogeneous, they are coset spaces:
\begin{equation}
\HH=\frac{\SO(1,4)}{\SO(4)}\;\;\quad \text{and} \quad
\Bergman=\frac{\SU(2,1)}{\Urm(2)}\;\;.
\end{equation}
As we will see explicitly in the next section, QTN 
smoothly interpolates between these two manifolds\footnote{By ``interpolation'' we mean
that there is a continuous range of QTN metrics (labelled as we will see by the ``nut charge''), 
two specific examples of which correspond to $\HH$ and $\Bergman$. So this is different from
  RG flows in AdS/CFT (see e.g. \cite{DHokerFreedman02} for 
orientation), where one interpolates between different solutions of some gauged supergravity
 (usually corresponding to UV and IR regimes in the boundary theory) by varying the radial parameter 
 which is interpreted as an energy scale. However, as we will see in section 8, one could 
perhaps consider the deformation of the boundary metric as a sort of \emph{marginal} flow.}. 
Note that there is no analogous interpolation via a smooth manifold 
between the compact versions of these homogeneous spaces, i.e. $\Srm^4$ and $\Cset\Prm^2$.

We have already mentioned one way to derive QTN, as the ``filling in'' manifold of
the Berger sphere (\ref{Bergersphere}).
QTN can also be seen to be a special case of the (Anti--) de Sitter--Taub--NUT/Bolt 
family of metrics 
that appears e.g. in \cite{Eguchietal80}. In the notation of \cite{Chamblinetal99} this
more general metric is
\begin{equation} \label{AdSTN}
\diff s^2=V(r)(\diff \tau+2n\cos\theta\diff\varphi)^2+\frac{\diff r^2}{V(r)}
+(r^2-n^2)(\diff\theta^2+\sin^2\theta\diff\varphi^2)
\end{equation}
with $V(r)=\left[(r^2+n^2)^2-2Mr+k^{2}(r^4-6n^2r^2-3n^4)\right]/(r^2-n^2)$. Apart from
the cosmological constant ($\grL=-3k^2$) there are two parameters, $M$ and $n$, which
correspond to a ``mass'' and ``nut charge'' respectively.
As explained thoroughly in \cite{Chamblinetal99}, to obtain 
a nut, i.e. a regular zero--dimensional fixed point set, we require that $V(r=n)=0$, so 
$M$ and $n$ are related by
\begin{equation} \label{AdSTNcond}
M=n-4 n^3k^2\;.
\end{equation}
A different possible choice for $M$ would lead to a regular two--dimensional 
fixed point set (a ``bolt'').  It is the special case of (\ref{AdSTNcond}) 
(thus the solution without bolts) that gives the QTN manifold. (The precise relation
to the metric found by Pedersen is given in appendix A.)

QTN has also been derived by Galicki \cite{Galicki87a,Galicki91} using the quaternionic 
K\"ahler quotient (described in detail in \cite{Galicki86,Galicki87b,GalickiLawson88})
 by considering a non--compact $\Rset$ action on the 8--dimensional
quaternionic projective ball {$\Hset \mathrm{H}^2=P(\Hset^{1,1}\times\Hset^1)$}. 
Finally, it has also been constructed by Ivanov and Valent \cite{IvanovValent98}
using harmonic superspace methods \cite{Galperinetal84,Galperinetal94}. (Actually the
term ``quaternionic Taub--NUT'' seems to have first appeared in \cite{IvanovValent98}.)

One application of QTN in the literature has been as the sigma--model manifold for scalars
coupled to $\Ncal=2$ supergravity. (Scalar couplings in $\Ncal=2$  supergravity are known 
to be parametrised by negative cosmological constant quaternionic K\"ahler manifolds
\cite{BaggerWitten83}.) It is in this context that QTN (and its higher dimensional
generalisations) was derived by Galicki. More recently, Behrndt and Dall'Agata 
\cite{BehrndtDallAgata02} considered 
abelian gaugings of its isometries and found domain wall solutions for
$d=5, \Ncal=2$ supergravity.  

 As we mentioned, QTN has already found some uses in an AdS/CFT
context (\cite{Hawkingetal99,Chamblinetal99,Emparanetal99,Mann99}). In 
\cite{Hawkingetal99,Chamblinetal99} it was used as a background for the
calculation of the gravitational action and entropy for the AdS--Taub--Bolt
manifold that has the same behaviour at infinity. In \cite{Emparanetal99} and \cite{Mann99} 
the action and entropy of AdS--Taub--NUT were calculated in an intrinsic 
way using the counterterm subtraction method of 
\cite{HenningsonSkenderis98,BalasubramanianKraus99b}. The result is somewhat unsettling in
that the gravitational entropy turns out to be negative, which is probably linked to problems
with the lorentzian interpretation \cite{Taylor--Robinson00}. 
Perhaps learning more about the dual CFT will help to understand this issue better.

\section{The Pedersen Metric for Quaternionic Taub--NUT}

The metric we will use to describe the quaternionic Taub--NUT manifold
is a slight rewriting of the one first written down by 
Pedersen \cite{Pedersen86}. Actually, in \cite{Pedersen86} there are two
metrics, which correspond to an oblate and prolate squashing of the boundary sphere 
\footnote{We recall that ``oblate squashing'' corresponds to flattening the sphere
at the poles, while ``prolate squashing'' means elongating the sphere at the poles.}.
We will treat them in turn.

\subsection{The Oblate case}

The form of the Pedersen metric that we will use is:
\begin{equation} \label{Pedersenoriginal}
g_{\mn}^{(m,k)}:\quad \diff s^2=\frac{4}{(1-k^2r^2)^2}\left[\frac{1+m^2 r^2}{1+m^2 k^2r^4}
\diff r^2+r^2 (1+m^2 r^2)(\sone^2+\stwo^2)+\frac{r^2(1+m^2 k^2 r^4)}
{1+m^2 r^2}\sthree^2\right]\;.
\end{equation}
Here $\sone,\stwo,\sthree$ are the usual $\SU(2)$ one--forms, defined by
\begin{equation}
\begin{split}
\sone&=\half (\cos\psi\diff\theta+\sin\psi\sin\theta\diff\varphi)\;,\\
\stwo&=\half (-\sin\psi\diff\theta+\cos\psi\sin\theta\diff\varphi)\;,\\
\sthree&=\half(\diff\psi+\cos\theta\diff\varphi)\;.
\end{split}
\end{equation}
(where $0\leq\theta\leq\pi,0\leq\varphi\leq2\pi,0\leq\psi\leq 4\pi$).
 With this convention they satisfy the cyclic relation:
\begin{equation}
\diff \sone=-2\stwo\wedge \sthree\;, \text{etc.}
\end{equation}

We have modified the expression in \cite{Pedersen86} by explicitly including
the cosmological constant, which can be seen to be $\Lambda=-3 k^2$. The 
parameter $m$ will soon be seen to correspond to the nut charge. So the metric
depends on two (real) parameters, $m$ and $k$, both with dimension $[L]^{-1}$, 
which can be varied independently. We can restrict to positive values of 
$m$ and $k$ without loss of generality \cite{Hitchin95}.    
The scalar curvature can be  seen to be $\Rcal=-12 k^2$. One can also check that
the Weyl tensor is anti--self--dual\footnote{In \cite{Pedersen86} the Weyl tensor turns out
to be self--dual due to a different convention for the sigmas 
($\diff\sigma_i=\sum\epsilon_{ijk}\sigma_j\wedge\sigma_k$).}. The isometry group is $\SU(2)\times\Urm(1)$
for generic values of $m,k$.

This metric is complete within the ball $r<1/k$ for all values of $m$
\cite{Hitchin95}\footnote{The singularities of quaternionic Taub--NUT
in various parameter ranges have also been considered in \cite{BehrndtDallAgata02}.}.
We will thus restrict $r$ to lie within the ball, with the boundary given by
$r\ra 1/k$. There the metric blows up, inducing the conformal structure:
\begin{equation}
\sone^2+\stwo^2+\frac{1}{1+m^2/k^2}\sthree^2\;.
\end{equation}
So this corresponds to an oblate squashing of the three sphere, since
the parameter $\grl$ in (\ref{Bergersphere}) is less than one.

\paragraph{The $k=0$ limit: Euclidean Taub--NUT}

Taking the cosmological constant $\Lambda=-3k^2$ to zero, while keeping the
nut charge nonzero, we obtain a Ricci--flat metric
\begin{equation} \label{ETN}
g_\mn^{(m,k=0)}:\quad \diff s^2=(1+m^2r^2)
[\diff r^2+r^2(\sone^2+\stwo^2)]
+\frac{ r^2}{1+m^2r^2}\sthree^2
\end{equation}
which is the metric on the euclidean Taub--NUT ALF space, in the form suitable for generalisation
to the multi--Taub--NUT metrics \cite{Hawking77,Eguchietal80}. We see that $m$ 
is indeed related to the nut charge (see appendix A for the precise relation).

\subsection{The Prolate case}

In the metric (\ref{Pedersenoriginal}) we assumed $m^2>0$, but it turns out that
this restriction is unnecessary \cite{Pedersen86}. Indeed, we can analytically
continue the parameter $m$ as follows\footnote{There is a deeper meaning to this analytic 
continuation which will become clear in section (\ref{More}).}: 
\begin{equation} \label{continuation}
m\ra i\mu\;.
\end{equation}
In this way we obtain the metric:
\begin{equation} \label{Pedersen}
g_\mn^{(\mu,k)}:\quad \diff s^2=\frac{4}{(1-k^2r^2)^2}\left[\frac{1-\mu^2 r^2}{1-\mu^2 k^2r^4}
\diff r^2+r^2 (1-\mu^2 r^2)(\sone^2+\stwo^2)+\frac{r^2(1-\mu^2 k^2 r^4)}
{1-\mu^2 r^2}\sthree^2\right]\;.
\end{equation}
This metric is complete within the ball $r<1/k$ as long as the range
 of $\mu$  is $(0,k)$ \cite{Hitchin95}. 
For $\mu>k$ the metric develops a true singularity, as can be seen
e.g. by the calculation of curvature invariants:
\begin{equation}
\Rcal^{\mn\kappa\grl}\Rcal_{\mn\kappa\grl}=24+24\frac{(1-k^2r^2)^2}{(1-\mu^2r^2)^2}\;.
\end{equation}
Accordingly, we will from now on restrict $\mu$ to lie in the range $(0,k)$, and again
$r$ to lie within the ball.
As for the induced conformal structure at the boundary ($r=1/k$), we find
\begin{equation} \label{prolate}
\sone^2+\stwo^2+\frac{1}{1-\mu^2/k^2}\sthree^2
\end{equation}
This gives a prolate squashing of the three--sphere, since now
$\grl$ in (\ref{Bergersphere}) is greater than one. 
\paragraph{The boundary metric}

By a choice of conformal scale, we can define a metric on the boundary 
that represents the conformal structure (\ref{prolate}). One possible choice is:
\begin{equation} 
h_{ij}^{(\mu,k)}: \quad \diff s^2=(1-\mu^2/k^2) (\sone^2+\stwo^2)+\sthree^2
\end{equation}
Here $i,j=1,2,3$. Recall that the value $\mu=0$  corresponds to the round 
($\SU(2)\times\SU(2)$ invariant) metric on $\Srm^3$. Away from that value, the isometry group 
is $\SU(2)\times\Urm(1)$, the same as the bulk manifold.
The scalar curvature of $h_{ij}^{(\mu,k)}$ is found to be:
\begin{equation}
\Rcal^{(3)}=\frac{3k^2-4\mu^2}{2(1-\mu^2/k^2)^2} 
\end{equation} 
As mentioned in the introduction, the scalar curvature becomes zero and
then negative as we increase $\mu^2$ from zero to its limiting value of $k^2$. 
For the value $\mu^2=\frac{3}{4}k^2$ we obtain $\Rcal^{(3)}=0$. As we approach
$k^2$ the (negative) scalar curvature blows up, indicating that the boundary metric becomes
degenerate. 

Returning to (\ref{Pedersen}), we note two limiting cases of special interest. 
\paragraph{The $\mu=0$ limit: Euclidean AdS$_4$ ($\HH$)}

 Taking $\mu=0$ in (\ref{Pedersen}) we obtain 
\begin{equation} \label{HH1}
g_{\mn}^{(\mu=0,k)}: \diff s^2=\frac{4}{(1-k^2r^2)^2}\left[\diff r^2
+r^2(\sone^2+\stwo^2+\sthree^2)\right]
\end{equation}
which is clearly the metric on $\HH$ in polar coordinates. (See 
appendix \ref{Coords} for the relation to the more commonly used upper half plane
metric on $\HH$.) The isometry group is now enhanced to $\SO(1,4)$, which will become the
conformal group of the boundary. As discussed, the boundary of $\HH$ is simply
$\Srm^3$ with the round metric, 
\begin{equation}
h_{ij}^{(0,k)}=\sone^2+\stwo^2+\sthree^2\;\;.
\end{equation}
The Weyl tensor of $\HH$ vanishes, corresponding to the fact that
it (and its boundary $\Srm^3$) are conformally flat.
\paragraph{The $\mu=k$ limit: The Bergman space}

Taking the $\mu=k$ limit of (\ref{Pedersen}), we obtain
\begin{equation} \label{Berg}
g_\mn^{(\mu=k,k)}:\;
\diff s^2=\frac{4}{(1-k^2r^2)^2}\left[\frac{\diff r^2}{1+k^2r^2}+
r^2(1+k^2r^2)\sthree^2\right]+\frac{4r^2(\sone^2+\stwo^2)}{1-k^2r^2}\;\;.
\end{equation}
This is the metric on the coset space $\Bergman=\SU(2,1)/\Urm(2)$, albeit in 
a slightly nonstandard form.
By performing the coordinate change
\begin{equation}
r^2=\frac{\rho^2}{2-k^2\rho^2}
\end{equation}
we obtain the more usual Bergman (pseudo--Fubini--Study) metric for $\Bergman$:
\begin{equation} \label{Bergmanmetric}
g_\mn^{(k,k)}:\quad
\diff s^2=2\frac{\diff \rho^2 +\rho^2\sthree^2}{(1-k^2\rho^2)^2}+2\frac{\rho^2(\sone^2+\stwo^2)}{(1-k^2\rho^2)}\,.
\end{equation}
We will thus loosely refer to the space $\Bergman$ as the ``Bergman space''. The isometry
group is now $\SU(2,1)$, and (according to \cite{Britto--Pacumioetal99}) becomes the conformal 
group of the boundary.

Note that the Bergman space is a K\"ahler manifold. So in this limit, our quaternionic 
K\"ahler metric actually becomes K\"ahler. As for the induced boundary metric, it is clear 
from (\ref{Berg}) that as one approaches the boundary ($r\ra 1/k$) the coefficient of 
$\sthree$ blows up faster that the others, and so the boundary metric 
degenerates\footnote{Technically, 
since the metric is K\"ahler, what extends to the boundary is not a conformal structure,
but a complex structure, and at the boundary we are left with a CR structure (which is generically
a degenerate complex structure) \cite{Hitchin95}.}. Despite this degeneracy, the authors of 
\cite{Britto--Pacumioetal99} showed that one could obtain meaningful results for 
correlation functions in the boundary theory.

\section{Propagators for the known cases}

 In trying to describe the boundary conformal field theory, a fundamental 
quantity is the bulk--to--boundary propagator. This parametrises the
coupling of a field in the bulk to its dual operator in the boundary theory,
and is essential in calculating boundary correlation functions. Although
in this analysis we are interested only in scalar two--point functions,
for which the bulk--to--boundary propagator is sufficient, we find it easier
and more intuitive to deal with the bulk--to--bulk propagators, from
which we can obtain the bulk--to--boundary ones via a limiting
process.   

 We note the laplacian for the oblate Pedersen metric (\ref{Pedersenoriginal}). 
(The corresponding
laplacian for (\ref{Pedersen}) can be found through $m=i\mu$): 
\begin{equation} \label{laplacianone}
\begin{split}
\nabla^2&=\frac{1}{\sqrt{g}}\partial_\mu\sqrt{g}g^\mn\partial_\nu\\
\\
&=\frac{(1-k^2r^2)^2}{(1+m^2 r^2)}
\left[\frac{1}{4}\left(1+m^2k^2r^4\right)\p_{rr}
 +\frac{1}{4r(1-k^2r^2)}\left(3+k^2r^2+7m^2k^2r^4-3m^2 k^4r^6\right)\p_r\right.\\
\\
&\left.+\frac{1}{r^2}\left\{\p_{\theta\theta}+\cot\theta\p_\theta
+\csc^2\theta\left[\p_{\ff\ff}-2\cos\theta\p_{\ff\psi}
+\left(\sin^2\theta\frac{(1+m^2r^2)^2}{1+m^2k^2r^4}+\cos^2\theta\right)
\p_{\psi\psi}\right]\right\}\right]\;.\\
\end{split}
\end{equation}

\comment{
\begin{equation} \label{laplacianone}
\begin{split}
\nabla^2&=\frac{1}{\sqrt{g}}\partial_\mu\sqrt{g}g^\mn\partial_\nu\\
\\
&=\frac{(1-k^2r^2)^2}{(1-\mu^2 r^2)}
\left[\frac{1}{4}\left(1-\mu^2k^2r^4\right)\p_{rr}
 +\frac{1}{4r(1-k^2r^2)}\left(3+k^2r^2-7\mu^2k^2r^4+3\mu^2 k^4r^6\right)\p_r\right.\\
\\
&\left.+\frac{1}{r^2}\left\{\p_{\theta\theta}+\cot\theta\p_\theta
+\csc^2\theta\left[\p_{\ff\ff}-2\cos\theta\p_{\ff\psi}
+\left(\sin^2\theta\frac{(1-\mu^2r^2)^2}{1-\mu^2k^2r^4}+\cos^2\theta\right)
\p_{\psi\psi}\right]\right\}\right]\\
\end{split}
\end{equation}
}

In general, a coordinate space Green's function $G(x_r,x_s)$ in four dimensions
will depend on eight variables. 
Following \cite{Britto--Pacumioetal99}, we use the $\SU(2)\times\Urm(1)$
isometry group to restrict propagators to depend on the following combinations of 
the angular parameters $\tr,\fr,\psi_r,\ts,\fs,
\psi_s$:\footnote{In \cite{Britto--Pacumioetal99}, $\SU(2)\times\Urm(1)$ appears as
the isometry group of the boundary of the Bergman space, while the bulk isometry
group is $\SU(2,1)$. In the more general QTN case both the bulk and the boundary 
have $\SU(2)\times\Urm(1)$ isometry group, and the same arguments apply.}
\begin{equation} \label{U}
U=\cos\frac{\tr}{2}\cos\frac{\ts}{2}
e^{\frac{i}{2}(\psi_s-\psi_r+\fs-\fr)}+
\sin\frac{\tr}{2}\sin\frac{\ts}{2}
e^{\frac{i}{2}(\psi_s-\psi_r-\fs+\fr)} \;\;.
\end{equation}
and its conjugate
\begin{equation} \label{Ubar}
\bar{U}=\cos\frac{\tr}{2}\cos\frac{\ts}{2}
e^{-\frac{i}{2}(\psi_s-\psi_r+\fs-\fr)}+
\sin\frac{\tr}{2}\sin\frac{\ts}{2}
e^{-\frac{i}{2}(\psi_s-\psi_r-\fs+\fr)} \;\;.
\end{equation}
The Green's function should be symmetric in $U$ and $\bar{U}$, so we find 
it convenient to introduce the two combinations\footnote{$\vone/2$ is called 
$\mu$ in \cite{Page79}.}:
\begin{equation} \label{vone}
\vone:=U+\bar{U}=2\cos\frac{\tr}{2}\cos\frac{\ts}{2}\cos\half(\psi_r+\fr-\psi_s-\fs)
+2\sin\frac{\tr}{2}\sin\frac{\ts}{2}\cos\half(\psi_r-\fr-\psi_s+\fs)
\end{equation}
and
\begin{equation} \label{vtwo}
\vtwo:=U\bar{U}=\half\{1+\cos\theta_r\cos\theta_s+
\sin\theta_r\sin\theta_s\cos(\varphi_r-\varphi_s)\}\;\;.
\end{equation} 
So from now on we will treat the bulk propagators as functions of $r,s,\vone,\vtwo$. That
is, we have used the isometries to reduce the number of variables from eight to four, and
by doing so we have exhausted the available symmetries of the metric\footnote{Note that, unlike
the definition of \cite{Britto--Pacumioetal99}, our $U,\bar{U}$ are pure angular variables.
This means that $\vone,\vtwo$ will always appear in Green's functions in the combinations
 $rs\vone,r^2s^2\vtwo$.}. In these new variables, $x_r=x_s$ translates to $\{r=s,\vone=2,\vtwo=1\}$. 

It is now useful to rewrite the laplacian (\ref{laplacianone}) in terms of $\vone$ and $\vtwo$:
\begin{equation} \label{laplacian}
\begin{split}
\nabla^2=\frac{(1-k^2r^2)^2}{(1+m^2 r^2)}
&\left[\frac{1}{4}\left(1+m^2 k^2r^4\right)\p_{rr}
 +\frac{1}{4r(1-k^2r^2)}\left(3+k^2r^2+7m^2 k^2 r^4-3m^2 k^4 r^6\right)\p_r\right.\\
\\
&+\frac{1}{r^2}\left\{
-1/4\vone\frac{(3+2m^2 r^2+2m^2 k^2 r^4 +m^4 r^4)}{1+m^2k^2r^4}\p_{\vone}\right.\\
&\left.+\frac{1+m^2 k^2 r^4-1/4\vone^2(1+m^2 r^2)^2-m^2 r^2 \vtwo
(-2+k^2 r^2-m^2 r^2)}{1+m^2k^2r^4}\p_{\vone\vone}\right.\\
&\left.\left.+(1-\vtwo) \vone \p_{\vone\vtwo}+(1-\vtwo)\vtwo\p_{\vtwo\vtwo}-(2\vtwo-1)\p_{\vtwo}
\right\}\right]\\
\end{split}
\end{equation}

\comment{
\begin{equation} \label{laplacian}
\begin{split}
\nabla^2=\frac{(1-k^2r^2)^2}{(1-\mu^2 r^2)}
&\left[\frac{1}{4}\left(1-\mu^2 k^2r^4\right)\p_{rr}
 +\frac{1}{4r(1-k^2r^2)}\left(3+k^2r^2-7\mu^2 k^2 r^4+3\mu^2 k^4 r^6\right)\p_r\right.\\
\\
&+\frac{1}{r^2}\left\{
-1/4\vone\frac{(3-2\mu^2 r^2-2\mu^2 k^2 r^4 +\mu^4 r^4)}{1-\mu^2k^2r^4}\p_{\vone}\right.\\
&\left.+\frac{1-\mu^2 k^2 r^4-1/4\vone^2(1-\mu^4 r^4-2\mu^2 r^2)+\mu^2 r^2 \vtwo
(-2+k^2 r^2+\mu^2 r^2)}{1-\mu^2k^2r^4}\p_{\vone\vone}\right.\\
&\left.\left.+(1-\vtwo) \vone \p_{\vone\vtwo}+(1-\vtwo)\vtwo\p_{\vtwo\vtwo}-(2\vtwo-1)\p_{\vtwo}
\right\}\right]\\
\end{split}
\end{equation}
}

In the following sections we give the known Green's functions for the various limiting
cases of this laplacian (and the one related to it by $m=i\mu$). 
We use the notation $G^{(\mu,k)}$ or $G^{(m,k)}$ for the Green's functions corresponding
to  the Pedersen metrics (\ref{Pedersen}) and (\ref{Pedersenoriginal}) respectively. 
We will discuss two common possibilities for scalars propagating in a curved spacetime, 
conformal and minimal coupling. Note that since the scalar curvature of QTN is $\Rcal=-12k^2$,
the conformally coupled laplacian is $\nabla^2-\frac{1}{6}\Rcal=\nabla^2+2k^2$.

\subsection{Green's Functions for $\HH$}

In this section we review a few well--known facts for propagators on euclidean AdS$_4$. All
this is completely standard material 
(\cite{BurgessLutken85,Witten98,MuckViswanathan98,Freedmanetal98,DHokeretal99a,DHokeretal99b,DHokeretal99c}, 
see also \cite{DHokerFreedman02} for a review), 
and is only included for easy comparison with
other limiting cases of QTN, and to familiarise the reader with our slightly
non--standard notation. (A sketch of how one converts to the usual expressions
for bulk--to--bulk and bulk--to--boundary propagators can be found in 
appendix B.) 

 One can easily invert the (massive) laplacian equation for $\HH$ 
 by assuming that the Green's function is a 
function only of the chordal distance $u_0:=u^{(0,k)}$, where 
(e.g. \cite{DHokeretal99c}). 
\begin{equation}
u^{(0,k)}=\frac{2k^2(r^2+s^2-rs\vone)}{(1-k^2r^2)(1-k^2s^2)}
\end{equation}
So the chordal distance allows us to express the $\mu=0$ limit of the laplacian 
(\ref{laplacian}) in terms of only one variable:
\begin{equation} \label{hyperAdS}
\nabla^2-M^2=2k^2\left\{u_0\frac{u_0+2}{2}\partial_{u_0u_0}
+2(u_0+1)\p_{u_0}-\frac{M^2}{2k^2}\right\}
\end{equation}

\comment{
Then the general solution of $(\nabla^2-M^2)G(u_0)=0$ is 
\begin{equation}
G^{0,k}=c_1 F(\dd_+,\dd_-;2;1+\frac{u_0}{2})+c_2 F(\dd_+,\dd_-;2;1+\frac{u_0}{2})
\int{\frac{\diff u_0}{u_0^2(u_0+2)^2 F(\dd_+,\dd_-;2;1+u_0/2)}}\;\;,
\end{equation}
where $F(a,b;c;x)$ is the hypergeometric function and 
\begin{equation}
\dd_\pm=\frac{3}{2}\pm\sqrt{\frac{9}{4}+\frac{M^2}{k^2}}\;\;.
\end{equation}

In general only one of these solutions (or a linear combination)  will be allowed 
by the boundary 
conditions we wish to impose.}

This is a hypergeometric equation whose solutions are known for arbitrary values of $M^2$. 
However, as mentioned, we shall concentrate on the following special cases:
\paragraph{Conformal coupling} Here $M^2=-2k^2$ and we obtain the two solutions
\begin{equation} 
\begin{split}
G_1^{(0,k)}&=C\cdot \frac{2k^2}{u_0}=C\cdot\frac{(1-k^2r^2)(1-k^2s^2)}{r^2+s^2-rs\vone}\\
G_2^{(0,k)}&=C\cdot \frac{4k^2}{u_0(u_0+2)}=C\cdot\frac{(1-k^2r^2)^2(1-k^2s^2)^2}
{(1+k^4r^2s^2-k^2rs\vone)(r^2+s^2-rs\vone)}\;.
\end{split}
\end{equation}
Notice that these propagators correspond to two different choices of 
boundary condition. This is the well known fact \cite{BreitenlohnerFreedman82a} 
that fields in AdS$_{d+1}$ in the mass range $-d^2/4<M^2/k^2<-d^2/4+1$
can have two possible quantisations. We will return to this point in section 8.
Both propagators reduce to the massless 
flat space Green's function as we take the cosmological constant to zero ($k\ra0$):
\begin{equation}
G_1^{(0,k)},G_2^{(0,k)}\longrightarrow G^{(0,0)}=C\cdot\frac{1}{r^2+s^2-rs\vone}\;. 
\end{equation}

The normalisation factors $C$ are known  for $\HH$, but since we are also 
interested in the more general QTN case we will not keep track of them. Instead,
we simply normalise all propagators (in this and the following sections) 
so that they give the same flat--space Green's function as $k\ra0$. 

From the bulk--to--bulk propagators one can find  bulk--to--boundary propagators by taking the
appropriate scaling limits (\cite{Giddings99,KlebanovWitten99b})\footnote{Of course, if we
had wanted to be careful with normalisation factors, we would have had to first 
define a proper Dirichlet problem at $r=1/k-\epsilon$ by analogy with 
\cite{Gubseretal98,MuckViswanathan99,Freedmanetal98}, define the bulk--to--boundary
propagators by normal derivatives of the bulk--to--bulk ones, and then take the regulator
to zero. Or we could add appropriate boundary terms to the action as done e.g. 
in \cite{KlebanovWitten99b}. Then for $\AdS_{d+1}$ we would obtain the relative factor $1/(2\Delta-d)$
between $G$ and $K$ ($\Delta$ is defined in section 8).}:
\begin{equation} \label{KHone}
K_1^{(0,k)}=\lim_{s\ra1/k}\left\{\frac{G_1^{(0,k)}}{(1-k^2s^2)}\right\}
=\frac{1-k^2r^2}{1+k^2r^2-kr\vone}
\end{equation}
and
\begin{equation} \label{KHtwo}
K_2^{(0,k)}=\lim_{s\ra1/k}\left\{\frac{G_2^{(0,k)}}{(1-k^2s^2)^2}\right\}
=\left(\frac{1-k^2r^2}{1+k^2r^2-kr\vone}
\right)^2\;.
\end{equation}
 
We have normalised these and all subsequent bulk--to--boundary propagators to give $1$ as
$r\ra0$.  
\paragraph{Minimal Coupling} Here $M^2=0$ and we have only one solution (the other is
a constant):
\begin{equation}
G_3^{(0,k)}=C\cdot
 k^2\left[-\ln\left(1+\frac{2}{u_0}\right)+\frac{2}{u_0}\frac{u_0+1}{u_0+2}\right]\;.
\end{equation}
Notice that $G_3^{(0,k)}$ also reduces to the massless 
flat space Green's function when $k\ra0$, as it should.
To obtain the bulk--to--boundary propagator, we need to take the limit
\begin{equation}
K_3^{(0,k)}=\lim_{s\ra1/k}\left\{\frac{G_3^{(0,k)}}{(1-k^2s^2)^3}\right\}
=\left(\frac{1-k^2r^2}{1+k^2r^2-kr\vone}\right)^3\;.
\end{equation}

\subsection{Green's Functions for $\widetilde{\Cset \Prm^2}$} \label{BergGreens}

In the case of the Bergman space, one can again simplify the laplacian 
equation considerably by assuming that the Green's function is a function
only of the chordal distace, which we denote by $u_1:=u^{(\mu=k,k)}$.  Expressions for the 
chordal distance and scalar Green's functions in the
case of $\Cset\Prm^2$ were found by Warner in \cite{Warner82}. We simply extend
these to the noncompact case by obvious changes in signs. After converting to our
coordinate system for $\Bergman$, we find
\begin{equation}
u_1=2k^2\frac{r^2+s^2+2k^2r^2s^2\vtwo-rs\sqrt{(1+k^2r^2)(1+k^2s^2)}\vone}{(1-k^2r^2)(1-k^2s^2)}\;.
\end{equation}

With this ansatz the laplacian becomes
\begin{equation} \label{hyperBerg}
\nabla^2-M^2=2k^2\left\{u_1(u_1+1)\p_{u_1u_1}+(2+3 u_1)\p_{u_1}-\frac{M^2}{2k^2}
\right\}\;.
\end{equation}
\comment{
Integrating $(\nabla^2-M^2/(2k^2))G(u_1)=0$, we find 
\begin{equation}
G^{(k,k)}=c_1\cdot F(\dd_+,\dd_-;2;-u_1)+c_2\cdot F(\dd_+,\dd_-;2;-u_1)
\int\frac{\diff u_1}{u_1^2(u_1+1)^2 F(\dd_+,\dd_-;2;-u_1)^2}
\end{equation}
where now 
\begin{equation}
\dd_\pm=1\pm\sqrt{1+\frac{M^2}{k^2}}\;.
\end{equation}
}
which is again a hypergeometric equation. As before we will restrict to the two 
special cases of conformally and minimally coupled scalars.
\paragraph{Conformal Coupling} Setting $M^2=-2k^2$ we again obtain two bulk--to--bulk propagators. 
The first one is easy to understand:
\begin{equation}
G_1^{(k,k)}=C\cdot\frac{2k^2}{u_1}=C\cdot\frac{(1-k^2r^2)(1-k^2s^2)}{r^2+s^2+2k^2r^2s^2\vtwo
-rs\sqrt{(1+k^2r^2)(1+k^2s^2)}\vone}\;.
\end{equation} 

This Green's function has a nice boundary behaviour, vanishing as $(1-k^2r^2)^1$ at the 
boundary (where $r\ra1/k$), like the corresponding mode in $\HH$. 
Taking the limit $k\ra0$, the Bergman space reduces to flat space and, as 
expected, we retrieve the flat space propagator as the $k\ra0$ limit of 
$G_1^{(k,k)}$. 
 
From the bulk--to--bulk propagator we find the bulk--to--boundary propagator (which is
essentially the Bergman kernel for the ball in $\Cset^2$) in the standard way
\begin{equation}
K_1^{(k,k)}=\lim_{s\ra1/k}\left\{\frac{G_1^{(k,k)}}{(1-k^2s^2)}\right\}
=\frac{1-k^2r^2}{1+k^2r^2+2k^2r^2\vtwo-\sqrt{2}kr\sqrt{1+k^2r^2}\vone}
\end{equation}

The second solution for conformal coupling is $\sim 1/u_1\cdot \ln(1+u_1)$, i.e. at the 
boundary it behaves as
$(1-k^2r^2)\ln(1-k^2r^2)$. This is very different from the corresponding mode in the $\HH$ case.
As we will see at the end of section 8, it might eventually become necessary to understand
this mode better.
\paragraph{Minimal Coupling} The solution for $M^2=0$ is
\begin{equation}
G_2^{(k,k)}= C\cdot k^2\left[-\ln\left(1+\frac{1}{u_1}\right)+\frac{1}{u_1}\right]
\end{equation}
One can check that this also reduces to the flat space propagator as $k\ra0$. 
To obtain the bulk--to--boundary propagator, we need to take the limit
\begin{equation}
K_2^{(k,k)}=\lim_{s\ra1/k}\left\{\frac{G_2^{(k,k)}}{(1-k^2s^2)^2}\right\} 
=\left(\frac{1-k^2r^2}{1+k^2r^2+2k^2r^2\vtwo-
\sqrt{2}kr\sqrt{1+k^2r^2}\vone}\right)^2\;\;.
\end{equation}
This bulk--to--boundary propagator is the same one that appears in 
\cite{Britto--Pacumioetal99} for minimal coupling (to see this one has to convert to the metric
(\ref{Bergmanmetric}) as explained in section 3).

\subsection{Green's Function for euclidean Taub--NUT}

Since euclidean Taub--NUT (ETN) is Ricci--flat, there is no distinction between the conformally
coupled and massless Green's function.  The massless scalar Green's function for 
ETN has been found by Page \cite{Page79}. In our parametrization, it is:
\begin{equation} \label{GreensTN}
G^{(m,0)}=
\frac{C}{2}\cdot\frac{2\cosh{(m^2 \mathrm{w})}-\frac{(r^2+s^2)}{\mathrm{w}}
\sinh{(m^2 \mathrm{w})}}
{(r^2+s^2)\cosh{(m^2 \mathrm{w})}-2\mathrm{w}\sinh{(m^2 \mathrm{w})}-rs\vone}\;\;,
\end{equation}
where 
\begin{equation} \label{w}
\mathrm{w}:=\half\sqrt{(r^2-s^2)^2-4r^2s^2(\vtwo-1)}\;\;.
\end{equation}
The euclidean Taub--NUT metric (\ref{ETN}) reduces to flat space by taking $m\ra0$. 
Taking this limit of $G^{(m,0)}$,  we again obtain the flat--space 
Green's function  $G^{(0,0)}$. ETN has zero cosmological constant, so we will 
not attempt to define bulk--to--boundary propagators since (for the moment) we are not
interested in defining a boundary theory.

\section{The Twistor space of the Pedersen metric}

After reviewing the Green's functions for various special cases, we are now ready
to consider the case of $\nabla^2-M^2$, where $\nabla^2$ is the full Pedersen laplacian
(\ref{laplacian}). 
The calculation we will perform is similar in spirit to the
one by Page in \cite{Page79}, where the Green's functions for scalar fields on
multi--Taub--NUT manifolds and ALE multi--instantons were produced. 

As in \cite{Page79}, the objective is to find functions of the coordinates
that vanish when two points are null separated. In \cite{Page79} these 
functions were found using Hamilton--Jacobi methods. We prefer to use
twistor methods in order to connect with the treatment of \cite{Pedersen86}. 
This discussion is rather long and the reader who is not interested in twistors
may prefer to skip to (\ref{eU}).

\subsection{More on the Pedersen metric} \label{More}

In preparation for the description of the twistor space, we briefly
review the main steps of the construction of the metric on QTN
by Pedersen in \cite{Pedersen86}.
Essentially the construction of \cite{Pedersen86} is similar to that of Gibbons and 
Hawking \cite{GibbonsHawking79} for the hyperk\"ahler case: One considers
a $\Urm(1)$ monopole over a 3-dimensional riemannian manifold $X^3$. This $\Urm(1)$ 
monopole is just a connection $A$ (the ``gauge potential'') on a principal 
$\Urm(1)$ bundle fibring over $X^3$ along with a function $V$ 
(the ``Higgs field'') that is related to the curvature of $A$ by
the Bogomolny equation
\begin{equation} \label{Bogomolny}
\diff A=-\star \diff V\;.
\end{equation}
With this choice of sign the conformal structure 
\begin{equation} \label{conformalstructure}
 V \diff s^2_{X^3}+V^{-1}(\diff \tau+A)^2\;,
\end{equation}
(where $\diff s_{X^3}^2$ is the metric on $X^3$) is anti--self--dual when $X^3$ has 
constant curvature. To obtain an Einstein space from this conformal structure,
one has to multiply by a suitable function $F(\chi)$. So given a choice
for $V$ satisfying (\ref{Bogomolny}) on $X^3$, and a corresponding choice for
$F(\chi)$, the metric
\begin{equation}
\diff s^2=F^2(\chi)(V\diff s_{X^3}^2+V^{-1}(\diff \tau+A)^2)
\end{equation}
will be Einstein. The flat case $X^3=\Rset^3$ is the 
Gibbons--Hawking ansatz, which leads to a zero scalar curvature 4--manifold 
(then $F(\chi)=1$). 
Pedersen's ansatz replaces $X^3$ by $\Srm^3$ and the 3--metric is chosen
to be $\diff s_{\Srm^3}^2=\diff\chi^2+4\sin^2\chi(\sigma_1^2+\sigma_2^2)$. A choice that
satisfies (\ref{Bogomolny}) is
 $(V,A)=(1+k/m\cot\chi,-k/m\cos\theta\diff\varphi)$ and after multiplying by a 
suitable conformal factor (and rescaling $\tau=-2k/m\cdot\psi$) one obtains an anti--self--dual 
Einstein  4--metric with negative cosmological constant \cite{Pedersen86}:
\begin{equation}
k^2\diff s^2=\frac{1}{(\cos\chi-\frac{k}{m}\sin\chi)^2}
\left[(1+\frac{k}{m}\cot\chi)\left\{\diff\chi^2+4\sin^2\chi(\sone^2+\stwo^2)\right\}
+\frac{4k^2\sthree^2}{m^2(1+\frac{k}{m}\cot\chi)}\right]\;.
\end{equation}

This is the ``trigonometric'' case in \cite{Hitchin95}. The coordinate change
\begin{equation}
\cos\chi=\frac{1}{\sqrt{1+m^2k^2r^4}}
\end{equation}
shows that it is equivalent to the original ``oblate'' Pedersen metric 
(\ref{Pedersenoriginal}).

There is another possibility that gives a negative cosmological constant, which
is to consider $X^3=H^3$, the three dimensional hyperbolic space, with metric
$\diff s_{H^3}^2=\diff\chi^2+4\sinh^2\chi(\sigma_1^2+\sigma_2^2)$.
 Then the Einstein 4--metric that arises is \cite{Hitchin95}:
\begin{equation}
k^2\diff s^2=\frac{-1}{(\cosh\chi-\frac{k}{\mu}\sinh\chi)^2}
\left[(1-\frac{k}{\mu}\coth\chi)\left\{\diff\chi^2+
4\sinh^2\chi(\sone^2+\stwo^2)\right\}
+\frac{4k^2/\mu^2\sthree^2}{(1-\frac{k}{\mu}\coth\chi)}\right].
\end{equation}

In this ``hyperbolic'' case we can perform the coordinate transformation 
\begin{equation}
\cosh\chi=\frac{1}{\sqrt{1-\mu^2k^2r^4}}
\end{equation}
to see that it is the same as the prolate Pedersen metric (\ref{Pedersen}). 
Thus the analytic continuation (\ref{continuation}) simply takes us from the trigonometric
case to the hyperbolic one.

\subsection{The twistor space}
\indent

 In the previous section we saw how to obtain an Einstein space with
anti--self--dual conformal structure via a  $\Urm(1)$ monopole over
a three--dimensional space of constant curvature. There is a beautiful
description of this situation in terms of twistor theory 
\cite{Hitchin82,JonesTod85}. Although all the material in this section can be 
found (in much greater detail!) in the literature, in the interest of 
completeness we will attempt to review the path leading to the twistor space 
 of our metric. Much of this and the following subsection is taken verbatim from 
\cite{Pedersen86}\footnote{We apologise in advance to any mathematician readers for what is 
certainly a naive and oversimplified exposition of this elegant but often
delicate construction.}. First we need to give a few definitions.

 The \emph{twistor space} of a self--dual\footnote{By self--dual
we mean that the Weyl tensor is self--dual. We switched to self--dual language for consistency
with the twistor literature, but the following discussion can be adapted  to the anti--self--dual
case simply by replacing ``self--dual'' with ``anti--self--dual'' everywhere.} four--dimensional 
Riemannian manifold $M$ is the projective bundle $Z=P(V_-)$ of anti--self--dual
spinors, and is a complex three--manifold \cite{Penrose76,Atiyahetal78,Hitchin79,Ward80}. 
The crucial property of $Z$ is
that its complex structure encodes the conformal structure of $M$. The way this works is,
roughly, the following: There is a holomorphic projection $Z\ra \Cset\Prm^1$, whose sections
are called \emph{twistor lines}. These lines (which are rational curves--copies of 
$\Cset\Prm^1$--that have normal bundle $\mathcal{O}(1)\oplus\mathcal{O}(1)$) belong to a four
complex--dimensional family, and their parameter space defines a complex 
self--dual conformal structure. Imposing reality under the real structure of $Z$ reduces
this to the real self--dual conformal structure of $M$. (This construction easily generalises to 
hyperk\"ahler and quaternionic K\"ahler manifolds in higher dimensions.)
 
 LeBrun \cite{LeBrun82} showed the following fact, which underlies Pedersen's construction: 
If ${N_C}$ is a three--complex dimensional manifold with holomorphic metric (with an
added property of being geodesically convex), then 
the space of unparametrised null geodesics of ${N_C}$ is the twistor space $Z$ of a 
four--complex--dimensional manifold $M_C$, which satisfies the self--dual Einstein equations with
negative cosmological constant. The 3--manifold  ${N_C}$ appears as the conformal infinity
(in the sense of Penrose) of the 4--manifold $M_C$. Now if $N_C$ is the complexification
of a real--analytic 3--manifold $N$, using the real structure of $Z$ we can find a real slice $M$ 
of $M_C$. In this way we can describe how a three (real) dimensional manifold becomes 
the conformal infinity of a four (real) dimensional self--dual Einstein space with negative
cosmological constant. 

 We now see that to apply the LeBrun construction to the Berger sphere (\ref{Bergersphere}) in
order to find the twistor space $Z$ of the filling--in manifold $M$, 
Pedersen had to somehow describe the space of unparametrised null geodesics of the Berger sphere.
 By using an analogy with the motion of a free rigid body around a fixed point, he was able 
to show that $Z$ can be described as a line bundle over plane sections of the quadric in 
$\Cset\Prm^3$. Using the notation $z_i$ for homogeneous coordinates on $\Cset \Prm^3$, 
this quadric is given by
\begin{equation} \label{quadric}
Q\;:\;z_1^2+z_2^2+z_3^2+\nu^2 z_4^2=0\;.
\end{equation}
The parameter $\nu$ is related to the squashing parameter $\grl$ in (\ref{Bergersphere}) 
by $\nu^2=\grl(1-\grl)$. By plane sections we mean the conics obtained by intersections of
the quadric with certain planes in $\Cset\Prm^3$. 
We will not go deeper into understanding the space of null geodesics, because there is an 
alternate description of $Z$ that will prove more useful.  

 To provide that description we need to introduce the concept of the \emph{minitwistor space} of
an Einstein--Weyl 3--manifold $X$ \cite{Hitchin82a,JonesTod85}\footnote{An Einstein--Weyl 
manifold with a conformal metric 
$g_{ij}$ satisfies $\Rcal_{(ij)}=\grL g_{ij}$, where $\Rcal_{ij}$ is the (possibly nonsymmetric)
Ricci tensor of a generically non--Levi--Civita connection \cite{JonesTod85}. 
We will only consider the special case of Einstein's equations, which in three 
dimensions imply constant curvature.}. This is a complex \emph{two}--manifold that contains 
certain
``special'' rational curves (again called twistor lines) with normal bundle $\mathcal{O}(2)$ 
which can be shown to form a three--parameter family. So the Einstein--Weyl manifold $X$ 
appears as the parameter space of these special curves. There is a way \cite{JonesTod85} to
pass from the twistor space $Z$ of a four--dimensional self--dual manifold $M$ to the minitwistor 
space 
of a three--dimensional Einstein--Weyl manifold $X$ that appears as the space of trajectories of a 
conformal motion on $M$, by suitably factoring $Z$ with a holomorphic vector field. This may
sound vague but in the cases we are interested in, the relation between twistor and minitwistor
spaces becomes very precise \cite{Pedersen86,Hitchin95}:

\emph{
The twistor space $Z$ of a $\Urm(1)$ monopole over a space $X^3$ of constant 
curvature is given by a holomorphic $\Cset^*$--bundle over the minitwistor
space of $X^3$, of degree zero on each twistor line of the minitwistor space.
}

To see the relevance of this, recall that in the previous section we showed how the 
Pedersen metric is constructed as
a $\Urm(1)$ monopole over $\Srm^3$ or $H^3$. So we now have a way to construct the twistor space
of the Pedersen metric. 
We will restrict to the oblate case for which $X^3$ is simply round $\Srm^3$. 
The minitwistor space of $\Srm^3$ turns out to be just \emph{the quadric
in $\Cset \Prm^3$}. This is of course not a coincidence since there is a close connection
between the monopole and the LeBrun descriptions. 

The quadric, then, plays a dual role: It parametrises the null geodesics of the squashed
three--sphere and also encodes, through its twistor lines, a \emph{round} three--sphere
$\Srm^3$. In the following we will use this second description to construct $Z$ and 
then, at the end, revert to the first one (since we are interested in the conformal
infinity of QTN and not in its description as a $\Urm(1)$ monopole).
The fact that the quadric (\ref{quadric}) somehow encodes a round sphere will become clearer
if we introduce affine coordinates. To do that, we need to set up patches. First, 
define the lines 
\begin{equation}
\begin{split}
l_1&: \; z_2+i z_3=0\;\wedge \;z_1+i\nu z_4=0\;, \\
l_2&: \; z_2+i z_3=0\;\wedge \;z_1-i\nu z_4=0\;, \\
m_1&: \; z_2-i z_3=0\;\wedge \;z_1+i\nu z_4=0\;, \\
m_2&: \; z_2-i z_3=0\;\wedge \;z_1-i\nu z_4=0\;.
\end{split}
\end{equation}

On the patch $U_1=Q\backslash (l_1\cup l_2)$ we will use the following 
coordinates:
\begin{equation}
\zeta=\frac{z_1+i\nu z_4}{z_2+i z_3}\;,\quad 
\eta=\frac{z_1-i\nu z_4}{z_2+i z_3}
\end{equation}
while on the other patch $U_2=Q\backslash (m_1\cup m_2)$ a valid coordinate
system is
\begin{equation}
\tilde{\zeta}=\frac{z_1-i\nu z_4}{z_2-i z_3}\;,\quad 
\tilde{\eta}=\frac{z_1+i\nu z_4}{z_2-i z_3}\;\;.
\end{equation}

Notice that the coordinates on the overlap of the two patches are related 
simply by $\tilde{\zeta}=-1/\zeta,\tilde{\eta}=-1/\eta$. Thus with this choice
of affine coordinates we have exhibited the quadric $Q$ as the product
$\Cset \Prm^1\times \Cset \Prm^1$. As stated above, the fact that $Q$ is the 
minitwistor space of $\Srm^3$
means that we can find special curves (the twistor lines) in $Q$ whose
parameter space is $\Srm^3$. These twistor lines turn out to be plane sections,
i.e. intersections of the quadric with hyperplanes in $\Cset \Prm^3$. As explained in 
\cite{Pedersen86} in affine coordinates this translates to the condition
\begin{equation} \label{planesections}
\eta=\frac{-\bar{b}+\bar{a}\zeta}{a+b\zeta}\;\;.
\end{equation}
We have introduced the complex parameters $a,b$ that satisfy 
$a\bar{a}+b\bar{b}=1$, i.e. they are coordinates on a round  $\Srm^3$. This is how
$\Srm^3$ appears as the parameter space of the twistor lines of its 
corresponding minitwistor space $Q$.

Having found the twistor lines of the quadric, the next step 
(according to the general description above) is to construct a line bundle $L$, 
trivial over these twistor lines.  This bundle will generate both the conformal 
structure we aim for, and a $U(1)$ monopole described by $(V,A)$ above.
Assume we have such a line bundle $L$, defined on a neighbourhood of a 
plane section of the quadric. The fact that $L$ is trivial over plane sections
means that if a section of $L$ is given by the holomorphic functions 
$(\sigma_1,\sigma_2)$ on the coordinate patches $U_1,U_2$ defined above, and
further if $\psi_{12}$ is the transition function with respect to 
$U_1,U_2$, then (on the plane sections)
\begin{equation} \label{trivial}
\sigma_1=\psi_{12}\sigma_2\;\;.
\end{equation}

Since we aim for a Higgs field $V=1+k/m\cot\chi$, Pedersen's idea is to 
construct $L$ as the direct product of (suitable powers of) the line
bundles $P$ and $T$ that give $V=i\cot\chi$ and $V=1$, respectively. 
The bundle $P$ is shown to have transition function
\begin{equation}
\psi_{12}^{(P)}(\zeta,\eta)=\frac{(\zeta-\eta)^2}{\zeta\eta}
\end{equation}
while the transition function for $T$ is just
\begin{equation}
\psi_{12}^{(T)}(\zeta,\eta)=\frac{\eta}{\zeta}\;\;.
\end{equation}
To introduce the parameter $m/k$ we can consider powers of the bundle $T$. So in the end
the transition function of the tensor product $L=T^{i\frac{m}{k}}\otimes P$ is
simply
\begin{equation} \label{transition}
\psi_{12}=(\psi_{12}^{(T)})^{i\frac{m}{k}}\cdot\psi_{12}^{(P)}\;\;.
\end{equation}
We have thus constructed a line bundle, trivial over plane sections of
the quadric, that naively gives the required monopole behaviour. 
Remarkably, Pedersen  actually proves that this simple guess is
correct, and thus (using the correspondence between the monopole description and the 
LeBrun construction) that

\emph{
The twistor space $Z$ of the Pedersen metric in 
the $\Srm^3$ (trigonometric) case (i.e. the case that gives a boundary conformal structure 
$\sone^2+\stwo^2+\grl\sthree^2$ with $\grl<1$) is the line bundle (minus the zero section)}
\begin{equation} 
L:=(T^{i\frac{m}{k}}\otimes P)\backslash 0
\end{equation}
\emph{
defined on a neighbourhood of a plane section of the quadric $Q$ in $\Cset \Prm^3$. 
}

\subsection{The real twistor lines}

In the previous section we obtained a description of the twistor space $Z$ of the
Pedersen metric in terms of a line bundle over plane sections of the quadric. 
We want to better understand the null structure of the Pedersen metric, so we need 
to examine the twistor lines of $Z$. 

First notice that the real structure on the quadric given, in terms of the homogeneous
coordinates, by {$\tau: (z_1,\ldots,z_4)\ra(\bar{z}_1,
\ldots,\bar{z}_4)$} takes the following form with respect to our affine
coordinates ($\zeta,\eta$):
\begin{equation} \label{realstructQ}
\tau\;:\;(\zeta,\eta)\ra(-1/\bar{\zeta},-1/\bar{\eta})\;.
\end{equation}
It is easy to see that the plane sections given by (\ref{planesections}) above
are real with respect to this real structure. 
As we saw in the previous section, these real plane sections parametrise an
$\Srm^3$ via the coordinates $a,\bar{a},b,\bar{b}$. The twistor lines of $Z$ 
should have one extra parameter, the fibre coordinate $\tau$ that describes
the real sections of $L$ over the real plane sections\footnote{We
expect no confusion to arise between the real structure and the fibre coordinate, 
both denoted by the letter $\tau$. We note also in passing that this is not the same $\tau$ as
the one in (\ref{conformalstructure}).}. The reason that 
we restrict to real sections is that we want $\tau$ to be real, to give a 
total of four real coordinates on QTN. 

To find the real sections of $L$, we need to extend the real structure
(\ref{realstructQ}) on the quadric to the full twistor space. This can
be done in various ways. Since $L$ doesn't contain the zero section, we
can choose the following real structure on the fibre: $\tau: z\ra \pm 1/\bar{z}$. 
Then the full real structure on $L$ is 
\begin{equation} \label{realstructure}
\tau: (\zeta,\eta,\sigma)\ra (-1/\bar{\zeta},-1/\bar{\eta},
\pm \bar{\psi}_{12}(\zeta,\eta)/\bar{\sigma})\;\;.
\end{equation}
Here $\psi_{12}$ is the transition function defined in (\ref{transition}).
We thus have all the necessary ingredients to describe the real section
$(\sigma_1,\sigma_2)$ on the real plane sections. 
First, substituting (\ref{planesections}) in  (\ref{trivial}) (with $\psi_{12}$ from 
(\ref{transition})) we have \cite{Pedersen86}
\begin{equation} \label{transf}
\psi_{12}=\left(\frac{-\bar{b}+\bar{a}\zeta}{\zeta(a+b\zeta)}\right)^{i\frac{m}{k}}
\cdot\left(\frac{b^2 (\zeta-\gra)^2(\zeta-\beta)^2}
{\zeta(a+b\zeta)(-\bar{b}+\bar{a}
\zeta)}\right)=\sigma_1\cdot\sigma_2^{-1}\;.
\end{equation}
Here we have introduced the roots $\gra$ and $\beta$ of the equation 
$b\zeta^2+(a-\bar{a})\zeta+\bar{b}=0$. We choose 
\begin{equation}
\gra=\frac{\bar{a}-a+\sqrt{d}}{2b}\;\;,\quad 
\beta=\frac{\bar{a}-a-\sqrt{d}}{2b}\;\;,
\end{equation}
with $d=(a-\bar{a})^2-4b\bar{b}$ the discriminant.
Second,  the requirement of reality under (\ref{realstructure}),  can be written
as
\begin{equation} \label{reality}
\sigma_1(\tau(\zeta))=\pm 1/\bar{\sigma}_2(\zeta)
\end{equation}
A choice for $\sigma_1,\sigma_2$ that satisfies both 
(\ref{transf}) and (\ref{reality}) is 
\begin{equation}
(\sigma_1,\sigma_2)=\left(i e^{i\tau} \frac{b}{\beta}\frac{(\zeta-\beta)^2}
{(a+b\zeta)^{im/k+1}},\frac{ie^{i\tau}}{b\beta}\frac{\zeta^{im/k+1}}{(\zeta-\gra)^2
(\bar{a}\zeta-\bar{b})^{im/k-1}}\right)\;\;.
\end{equation}

We finally have the full description of the real twistor lines over the 
real plane sections: They are given by (restricting to the  patch $U_1$ 
from now on):
\begin{equation}
\eta=\frac{-\bar{b}+\bar{a}\zeta}{a+b\zeta}
\end{equation}
and 
\begin{equation}
\sigma=i e^{i\tau}\frac{b}{\beta}\frac{(\zeta-\beta)^2}{(a+b\zeta)^{im/k+1}}\;.
\end{equation}

In the next section we will finally start building on Pedersen's work by 
using these twistor lines to construct the functions of null separation of the 
Pedersen metric.

\section{The functions of null separation}

Having found the twistor lines, we can now use Penrose's nonlinear graviton
construction \cite{Penrose76,Atiyahetal78,Hitchin79,Ward80} to derive the functions of 
null separation. We use the following well--known fact:{\it
Two points (say $x_r,x_s$) on the 4--dimensional manifold parametrised 
by $a,b,\bar{a},\bar{b},\tau$ are null separated if their corresponding twistor 
lines intersect.}
So the condition for $x_r$ and $x_s$ to be null separated is that the two equations
\begin{equation} \label{nulllines}
\text{(I):}\quad\eta_r\nceq\eta_s,\quad\quad \text{(II):}\quad\sigma_r\nceq\sigma_s
\end{equation}
have a common root. Here $\eta_r,\eta_s,\sigma_r,\sigma_s$ are the twistor lines
evaluated at the points $x_r$ and $x_s$, with corresponding coordinates
$a_r,b_r,\bar{a}_r,\bar{b}_r,\tau_r$ and $a_s,b_s,\bar{a}_s,\bar{b}_s,\tau_s$. 
(We will use the symbol $\nceq$ to denote equality on the null cone.) 
Clearly this description applies to the whole null cone, i.e. for arbitrary
separation of the two points $x_r$ and $x_s$. Often (as in \cite{Hitchin79} for instance)
 one considers the 
infinitesimal version of (\ref{nulllines}) (i.e. $x_r$ and $x_s$ close together), 
which is all that is required
to obtain the conformal structure (and eventually the Einstein manifold) 
described by $Z$. Here we are looking for functions that vanish on the whole
null cone, so we need to keep $x_r$ and $x_s$ at arbitrary separation.

The coordinates $a,\bar{a},b,\bar{b},\tau$ are the ones adapted to the
description of QTN as a $\Urm(1)$ bundle over $\Srm^3$ with 
the standard round metric. As discussed earlier, we need to convert these to the 
coordinates $r,\theta,\varphi,\psi$ that are more relevant to the description of
QTN as the manifold giving the Berger sphere as its conformal infinity.

The transformations that achieve this are \cite{Pedersen86}:
\begin{equation}
a=\frac{1+imkr^2\cos\theta}{\sqrt{1+m^2k^2r^4}}\quad,\quad
b=\frac{mkr^2\sin\theta}{\sqrt{1+m^2k^2r^4}}e^{i\varphi}
\end{equation}
and\footnote{Our variable $\psi$ differs by a sign from the one used by Pedersen. This
is required to match the conventions for the $\SU(2)$ one--forms: As mentioned earlier,
in \cite{Pedersen86}
$\diff\sigma_i=\sum\epsilon_{ijk}\sigma_j\wedge\sigma_k$, while we use 
$\diff\sigma_i=-\sum\epsilon_{ijk}\sigma_j\wedge\sigma_k$. Combined with a different 
choice for $\omega$ in \cite{Pedersen86}, p. 51 
(namely, $\omega=-\cos\theta\diff\varphi+\diff(\varphi-\chi)$), that is required by our
anti--self--dual ansatz (\ref{Bogomolny}), 
$\psi\ra-\psi$ takes $\sthree\ra-\sthree$.}  
\begin{equation} \label{psitau}
\tau=-\psi-\varphi+\arctan mkr^2
\end{equation}
Note also that the expression $\beta$ that appears in $\sigma$ is now
\begin{equation}
\beta=-i\frac{(1+\cos{\theta})}{\sin\theta}e^{-i\varphi}
\end{equation}

So we now begin the rather arduous process of converting the conditions (I) and
(II)  to expressions involving the points $x_r=(r,\theta_r,\varphi_r,\psi_r)$ and
$x_s=(s,\theta_s,\varphi_s,\psi_s)$. 
The first condition (I) is
\begin{equation}
\frac{-\bar{b}_r+\bar{a}_r\zeta}{a_r+b_r\zeta}\nceq
\frac{-\bar{b}_s+\bar{a}_s\zeta}{a_s+b_s\zeta}
\end{equation}
which has two solutions:
\begin{equation}
\zeta_\pm=\half
\frac{\bar{b}_rb_s+\bar{a}_sa_r-\bar{b}_sb_r-\bar{a}_ra_s\pm\sqrt{\Delta_0}}
{\bar{a}_rb_s-\bar{a}_sb_r}
\end{equation}
with the discriminant $\Delta_0$ given by
\begin{equation}
\begin{split}
\Delta_0=&\bar{b}_r^2b_s^2+\bar{a}_r^2a_s^2+\bar{b}_s^2b_r^2+\bar{a}_s^2a_r^2
-4(\bar{a}_rb_s\bar{b}_sa_r+\bar{a}_sb_r\bar{b}_ra_s)\\
&+2(\bar{b}_r b_s\bar{a}_r a_s-\bar{b}_rb_s\bar{b}_sb_r+\bar{b}_rb_s\bar{a}_sa_r
+\bar{a}_ra_s\bar{b}_sb_r-\bar{a}_ra_s\bar{a}_sa_r+\bar{b}_sb_r\bar{a}_sa_r)\;.
\end{split}
\end{equation}
Using the explicit expressions
\begin{equation}
\begin{split}
a_r&=\frac{1+imkr^2\cos\theta_r}{\sqrt{1+m^2k^2r^4}}\;\;,\quad\quad
a_s=\frac{1+imks^2\cos\theta_s}{\sqrt{1+m^2k^2s^4}}\;,\\
b_r&=\frac{mkr^2\sin\theta_re^{i\varphi_r}}{\sqrt{1+m^2k^2r^4}}\;\;\;\;,\quad\quad
b_s=\frac{mks^2\sin\theta_se^{i\varphi_s}}{\sqrt{1+m^2k^2s^4}}\;\,,
\end{split}
\end{equation}
and the definition of $\vtwo$ from (\ref{vtwo}), we rewrite $\Delta_0$ as 
\begin{equation}
\Delta_0=-\frac{4m^2k^2}{(1+m^2k^2r^4)(1+m^2k^2s^4)}{\Delta}
\end{equation}
where 
\begin{equation}
\Delta=(r^2-s^2)^2-4r^2s^2(\vtwo-1)-4m^2k^2r^4s^4\vtwo(\vtwo-1)\;\;.
\end{equation}

So the final expression for $\zp$ and $\zm$ is
\begin{equation}
\zeta_\pm=-i\frac{r^2\cos\tr-s^2\cos\ts-mkr^2s^2\sin\tr\sin\ts \sin(\fr-\fs)\pm\sqrt{\Delta}}{r^2\sin\tr e^{i\fr}(1-imks^2\cos\ts)-s^2\sin\ts e^{i\fs}(1-imkr^2\cos\tr)}\;.
\end{equation}
It is clear that $\zp\leftrightarrow\zm$ if we interchange $x_r$ with $x_s$. Also, it
can be checked that $\bar{\zeta}_+\zm=-1\;\;$.

Since condition (I) has two solutions, substituting
into (II) we obtain two conditions for the sigmas:
\begin{equation}
\sigma_r(\zp)\nceq\sigma_s(\zp)\;\;\text{and}\;\;
\sigma_r(\zm)\nceq\sigma_s(\zm)\;.
\end{equation}
So we are led to consider the following two functions that are equal to unity
when the points $x_r$ and $x_s$ are null separated:
 \begin{equation}
\Acal_{+}=\frac{\sigma_r(\zp)}{\sigma_s(\zp)}\nceq1\;\; \text{and}
\;\;\Acal_-=\frac{\sigma_r(\zm)}{\sigma_s(\zm)}\nceq1\;.
\end{equation}
To construct Green's functions it is useful to combine $\Acal_+$ and $\Acal_-$ so
as to obtain expressions that have definite symmetry 
properties under $x_r\leftrightarrow x_s$. To do that we
notice first of all that under $x_r\leftrightarrow x_s$, we have
\begin{equation}
\Acal_+\leftrightarrow\frac{1}{\Acal_-}\;\;.
\end{equation}
(To see this, consider that under $x_r\lra x_s$ we clearly have $a_r\lra a_s$,
$b_r\lra b_s$ and $\beta_r\lra \beta_s$ and as mentioned also 
$\zp\lra\zm$. Then $x_r\lra x_s$ takes $\sigma_r(\zp)\lra\sigma_s(\zm)$ and 
$\sigma_r(\zm)\lra\sigma_s(\zp)$.)

We thus define a symmetric (under $x_r\lra x_s$) function $\mathcal{U}(x_r,x_s)$ and 
an antisymmetric function $\mathcal{T}(x_r,x_s)$ by
\begin{equation}
e^{-4\mathcal{U}(x_r,x_s)}=\frac{\Acal_+}{\Acal_-}\;\;\text{and}\;\;
e^{-4i\mathcal{T}(x_r,x_s)}=\Acal_+\Acal_-\;.
\end{equation}
We choose the coefficients of $\mathcal{U}$ and $\mathcal{T}$ for consistency with Page 
\cite{Page79}. 
We will see that these two functions reduce to the ones found in that article in the
limit of zero cosmological constant. 

Now we have to calculate the functions $\mathcal{U}(x_r,x_s)$ and
$\mathcal{T}(x_r,x_s)$. We start by finding $\Acal_+/\Acal_-$:
\begin{equation}
\frac{\Acal_+}{\Acal_-}=\frac{\sigma_r(\zp)}{\sigma_s(\zp)}
\frac{\sigma_s(\zm)}{\sigma_r(\zm)}
=\left\{\frac{(\zp-\beta_r)(\zm-\beta_s)}{(\zp-\beta_s)(\zm-\beta_r)}\right\}^2
\left\{\frac{(a_s+b_s\zp)(a_r+b_r\zm)}{(a_r+b_r\zp)(a_s+b_s\zm)}
\right\}^{i\frac{m}{k}+1}\;.
\end{equation}
A rather long calculation gives the simple answer
\begin{equation}
\frac{(\zp-\beta_r)(\zm-\beta_s)}{(\zp-\beta_s)(\zm-\beta_r)}=
\frac{r^2+s^2-\sqrt{\Delta}-2imkr^2s^2\vtwo}{r^2+s^2+\sqrt{\Delta}-2imkr^2s^2\vtwo}
\end{equation}
and also 
\begin{equation}
\frac{(a_s+b_s\zp)(a_r+b_r\zm)}{(a_r+b_r\zp)(a_s+b_s\zm)}=
\frac{1+imk\sqrt{\Delta}+m^2k^2r^2s^2(2\vtwo-1)}{1-imk\sqrt{\Delta}+m^2k^2r^2s^2(2\vtwo-1)}\;.
\end{equation}
Combining these expressions, we obtain the final result for the symmetric combination
\begin{equation}
\frac{\Acal_+}{\Acal_-}=\left\{\frac{r^2+s^2-\sqrt{\Delta}-2imkr^2s^2\vtwo}
{r^2+s^2+\sqrt{\Delta}-2imkr^2s^2\vtwo}\right\}^2
\left\{\frac{1+imk\sqrt{\Delta}+m^2k^2r^2s^2(2\vtwo-1)}
{1-imk\sqrt{\Delta}+m^2k^2r^2s^2(2\vtwo-1)}\right\}^{1+i\frac{m}{k}}
\;.
\end{equation}

Now we turn to the antisymmetric combination $\Acal_+\Acal_-$.  We have:
\begin{equation} \label{ApAm}
\Acal_+\Acal_-
=\frac{\sigma_r(\zp)}{\sigma_s(\zp)}\frac{\sigma_r(\zm)}{\sigma_s(\zm)}
=\frac{e^{2i\tau_r}}{e^{2i\tau_s}}\frac{b_r^2}{b_s^2}\frac{\beta_s^2}{\beta_r^2}
\left\{\frac{(\zp-\beta_r)(\zm-\beta_r)}{(\zp-\beta_s)(\zm-\beta_s)}\right\}^2
\left\{\frac{(a_s+b_s\zp)(a_s+b_s\zm)}{(a_r+b_r\zp)(a_r+b_r\zm)}\right\}
^{1+\frac{im}{k}}\;.
\end{equation}
We will perform this calculation in steps. First we easily derive 
\begin{equation} \label{bb}
\frac{b_r^2}{b_s^2}=\frac{(1+m^2k^2s^4)r^4\sin^2\theta_r}{(1+m^2k^2r^4)s^4\sin^2\theta_s}
e^{2i(\varphi_r-\varphi_s)}
\end{equation}
and 
\begin{equation} \label{betabeta}
\frac{\beta_s^2}{\beta_r^2}=\frac{(1+\cos\theta_s)^2\sin^2\theta_r}
{(1+\cos\theta_r)^2\sin^2\theta_s}e^{2i(\varphi_r-\varphi_s)}\;.
\end{equation}

The remaining factors require a bit more work:
\begin{equation} 
\begin{split}
(\zp-\beta_r)&(\zm-\beta_r)=
s^2e^{-i\fr}(1+\cos\tr)(1-imkr^2)\\
&\times\frac{\left[\sin\ts(1-\cos\tr)e^{i(\fr-\fs)}-\sin\ts(1+\cos\tr)e^{-i(\fr-\fs)}
+2\cos\ts\sin\tr\right]}
{\sin^2\tr(r^2\sin\tr e^{i\fr}(1-imks^2\cos\ts)-s^2\cos\ts e^{i\fs}(1-imkr^2\cos\tr))}
\end{split}
\end{equation}
and
\begin{equation} 
\begin{split}
(\zp-\beta_s)&(\zm-\beta_s)=
r^2e^{-i\fs}(1+\cos\ts)(1-imks^2)\\
&\times\frac{\left[\sin\tr(1-\cos\ts)e^{-i(\fr-\fs)}-\sin\tr(1+\cos\ts)e^{i(\fr-\fs)}
+2\cos\tr\sin\ts\right]}
{\sin^2\ts(r^2\sin\tr e^{i\fr}(1-imks^2\cos\ts)-s^2\cos\ts e^{i\fs}(1-imkr^2\cos\tr))}
\;.
\end{split}
\end{equation}

We can simplify these expressions by observing that 
\begin{equation}
\begin{split}
\sin\ts(1-\cos\tr)&e^{i(\fr-\fs)}-\sin\ts(1+\cos\tr)e^{-i(\fr-\fs)}
+2\cos\ts\sin\tr=\\
&\quad4\left(\sin\frac{\tr}{2}\cos\frac{\ts}{2}e^{i\frac{\fr-\fs}{2}}
-\sin\frac{\ts}{2}\cos\frac{\tr}{2}e^{-i\frac{\fr-\fs}{2}}\right)\\
&\quad\times\left(\cos\frac{\tr}{2}\cos\frac{\ts}{2}e^{-i\frac{\fr-\fs}{2}}
+\sin\frac{\tr}{2}\sin\frac{\ts}{2}e^{i\frac{\fr-\fs}{2}}\right)\;,
\end{split}
\end{equation}
while
\begin{equation} 
\begin{split}
\sin\tr(1-\cos\ts)&e^{-i(\fr-\fs)}-\sin\tr(1+\cos\ts)e^{i(\fr-\fs)}
+2\cos\tr\sin\ts=\\
&\quad4\left(\sin\frac{\ts}{2}\cos\frac{\tr}{2}e^{-i\frac{\fr-\fs}{2}}
-\sin\frac{\tr}{2}\cos\frac{\ts}{2}e^{i\frac{\fr-\fs}{2}}\right)\\
&\quad\times\left(\cos\frac{\tr}{2}\cos\frac{\ts}{2}e^{i\frac{\fr-\fs}{2}}
+\sin\frac{\tr}{2}\sin\frac{\ts}{2}e^{-i\frac{\fr-\fs}{2}}\right)\;.
\end{split}
\end{equation}
The common factors cancel in the quotient, so we arrive at the simpler expression
\begin{equation}
\begin{split}
\frac{(\zp-\beta_r)(\zm-\beta_r)}{(\zp-\beta_s)(\zm-\beta_s)}=&
-\left(\frac{\cos\frac{\tr}{2}\cos\frac{\ts}{2}e^{-i\frac{\fr-\fs}{2}}
+\sin\frac{\tr}{2}\sin\frac{\ts}{2}e^{i\frac{\fr-\fs}{2}}}
{\cos\frac{\tr}{2}\cos\frac{\ts}{2}e^{i\frac{\fr-\fs}{2}}
+\sin\frac{\tr}{2}\sin\frac{\ts}{2}e^{-i\frac{\fr-\fs}{2}}}\right)\\
&\times \left(\frac{s^2e^{-i\fr}(1+\cos\tr)\sin^2\ts(1-imkr^2)}{r^2e^{-i\fs}(1+\cos\ts)\sin^2\tr
(1-imks^2)}\right)\;.\\
\end{split}
\end{equation}
Looking back at the definitions of $U$ and $\bar{U}$ from (\ref{U}) and (\ref{Ubar}), we see that
\begin{equation} \label{zpbeta}
\frac{(\zp-\beta_r)(\zm-\beta_r)}{(\zp-\beta_s)(\zm-\beta_s)}=
-\left(\frac{U}{\bar{U}}\right)e^{i(\psi_r-\psi_s)}e^{-i(\varphi_r-\varphi_s)}
 \left(\frac{s^2(1+\cos\tr)\sin^2\ts(1-imkr^2)}{r^2(1+\cos\ts)\sin^2\tr
(1-imks^2)}\right)\;.
\end{equation}

The last expression we need turns out to be
\begin{equation} \label{zpa}
\frac{(a_s+b_s\zp)(a_s+b_s\zm)}{(a_r+b_r\zp)(a_r+b_r\zm)}=1\;.
\end{equation}

Combining (\ref{bb}),(\ref{betabeta}),(\ref{zpbeta}) and (\ref{zpa}) and replacing
the variable $\tau$ by $\psi$ (using (\ref{psitau})) we see that (\ref{ApAm}) finally 
becomes
\begin{equation}
\Acal_+\Acal_-=\left(\frac{U}{\bar{U}}\right)^2\;.
\end{equation}

We thus conclude that the functions of $\mathcal{U},\mathcal{T}$ that reduce to 1 when the
points $x_r$ and $x_s$ are null separated are given by
\begin{equation} \label{eU}
e^{\mathcal{U}}=\left\{\frac{r^2+s^2+\sqrt{\Delta}-2imkr^2s^2\vtwo}{r^2+s^2-\sqrt{\Delta}-2imkr^2s^2\vtwo}\right\}^{\frac{1}{2}}
\left\{\frac{1-imk\sqrt{\Delta}+m^2k^2r^2s^2(2\vtwo-1)}{1+imk\sqrt{\Delta}+m^2k^2r^2s^2(2\vtwo-1)}\right\}^{\frac{1}{4}\left(1+i\frac{m}{k}\right)}
\end{equation}
and
\begin{equation}
e^{i\mathcal{T}}=\sqrt{\frac{\bar{U}}{U}}\;.
\end{equation}

Following  \cite{Page79}, we define the functions of null separation as
\begin{equation}
\mathcal{S}_\pm=1-e^{i\mathcal{T}\mp\mathcal{U}}\;.
\end{equation}
Our construction now assures that these functions vanish on the null cones.

\section{The Pedersen Green's function}

We now have functions of the coordinates $x_r,x_s$  that vanish 
when these points are on each other's null cones. Associating, as usual, the 
coordinate $\psi$ with euclidean time, we follow Page's argument that 
for fixed ``spatial'' coordinates (in this case,  $\vec{x}=(r,\theta,\varphi)$) 
the Green's function should only have 
simple poles in $\mathcal{S}_+$ and $\mathcal{S}_-$ at $\mathcal{S}_\pm=0$. 
Restricting to the simplest case, that of conformal coupling,  
we make the ansatz\footnote{We use the notation $G_1^{(m,k)}$ to emphasise that it
vanishes as $(1-k^2r^2)$ at the boundary, which we anticipate from the known
limits (section 4). We can also anticipate that $\Phi$ will diverge as $\sim 1/\sqrt{\Delta}$ as 
$\vec{x}_s$ is taken to be close to $\vec{x}_r$, since $\sinh\mathcal U\sim\sqrt{\Delta}$ in 
this limit. One
could turn this argument around, claiming that $G^{(m,k)}$ should not have a pole at 
$\vec{x}_r=\vec{x}_s$ if $\psi_r\neq\psi_s$, and thus try to write a more general ansatz for
other masses.}: 
\begin{equation} \label{ansatz}
G_1^{(m,k)}=\Phi(\vec{x}_r,\vec{x}_s)\left(\frac{1}{\mathcal{S}_+}-\frac{1}{\mathcal{S}_-}\right)
=\Phi(\vec{x}_r,\vec{x}_s)\left(\frac{\sinh \mathcal{U}}{\cosh \mathcal{U}-\cos\mathcal{T}}\right)\;.
\end{equation}
All that remains is to find the function $\Phi(\vec{x}_r,\vec{x}_s)$. To do this we can
integrate 
the laplacian (\ref{laplacianone}) over a cycle of $\psi$ ($0\leq\psi\leq4\pi$) to
obtain (after converting back to the remaining $\vtwo$ angular coordinate) 
\begin{equation}
\begin{split}
^{(3)}\nabla^2
=\frac{(1-k^2r^2)^2}{1+m^2r^2}&\left(\frac{1}{4}(1+m^2k^2r^4)\p_{rr}
+\frac{1}{4r(1-k^2r^2)}\left[3+k^2r^2+7m^2k^2r^4-3m^2k^4r^6\right]\p_r\right.\\
&+\left.\frac{1}{r^2}\left[\vtwo(1-\vtwo)\p_{\vtwo\vtwo}+(1-2\vtwo)\p_{\vtwo}\right]\right)\\
\end{split}
\end{equation} 
So for conformal coupling we look for a Green's function $^{(3)}G$ for the laplacian 
(on some ``auxiliary'' non--Einstein three--dimensional space)
\begin{equation} \label{threedlap}
(^{(3)}\nabla^2+2k^2) ^{(3)}G=0
\end{equation}
where $^{(3)}G$ is related to the Pedersen Green's function we are looking for by
\begin{equation} \label{int}
^{(3)}G(r,s,\vtwo)=\int_0^{4\pi}\diff\psi_r G_1^{(m,k)}(r,s,\vone,\vtwo)\;.
\end{equation}
Fortunately, it turns out that one can easily invert (\ref{threedlap}) to obtain
\begin{equation}
^{(3)}G(r,s,\vtwo)=\frac{(1-k^2 r^2)(1-k^2 s^2)}{\sqrt{\Delta}}
\end{equation}
where, as before, 
$\Delta=(r^2-s^2)^2-4r^2s^2(\vtwo-1)-4m^2k^2r^4s^4\vtwo(\vtwo-1)$.

The last step is to perform the integration in (\ref{int}). To do that we notice that 
$\cos\mathcal{T}=\cos((\psi_r-\psi_s)/2+\mathcal{S})$, where $\mathcal{S}=\arctan(\cos\half(\tr-\ts)/
\cos\half(\tr-\ts)\tan(\varphi_r-\varphi_s))$ and that (for an arbitrary $X$)
\begin{equation}
\int_0^{4\pi}\frac{\diff t}{a+b\cos(t/2+X)}=\frac{4\pi}{\sqrt{a^2-b^2}}\;.
\end{equation}
In our case $a=\cosh\mathcal{U}$ and $b=1$. So we can compute (\ref{int}) to be:
\begin{equation}
\int_0^{4\pi}\diff\psi_r G_1^{(m,k)}=4\pi\Phi(\vec{x}_r,\vec{x}_s)\frac{\sinh\mathcal{U}}
{\sqrt{\cosh^2\mathcal{U}-1}}
=4\pi\Phi(\vec{x}_r,\vec{x}_s)\;.
\end{equation}
Thus $4\pi\Phi(\vec{x}_r,\vec{x}_s)=\, ^{(3)}G(r,s,\vtwo)$. Introducing the notation 
\begin{equation}
\xi_{\pm}=(r^2+s^2\pm\sqrt{\Delta}-2imkr^2s^2\vtwo)(1\mp imk\sqrt{\Delta}+m^2k^2r^2s^2(2\vtwo-1))
^{\half(1+\frac{im}{k})}
\end{equation}
and 
\begin{equation}
\gamma=\left[(1+m^2k^2r^4)(1+m^2k^2s^4)\right]^{\frac{1}{4}\left(1+\frac{im}{k}\right)}
\sqrt{(1-imkr^2)(1-imks^2)}\;,
\end{equation}
and absorbing all normalisation factors into the usual coefficient $C$,
we conclude that the Green's function for a conformally coupled scalar
propagating on QTN is 
\begin{equation} \label{PedGreen}
G_1^{(m,k)}(r,s,\vone,\vtwo)
=C\cdot\frac{(1-k^2r^2)(1-k^2s^2)}{\sqrt{\Delta}}\frac{\xi_+-\xi_-}{\xi_++\xi_--2rs\gamma\vone}\;.
\end{equation}
This result can be verified by direct calculation\footnote{To show that $G_1^{(m,k)}$ is a Green's 
function for the laplacian (\ref{laplacian}) with conformal coupling, we find it easier to write 
it in the form (\ref{ansatz}) and keep the factors of $\cosh\mathcal{U}$ and $\sinh\mathcal{U}$
up to the end of the calculation. So for example if $G_1^{(m,k)}=\Phi\cdot F$, 
with $F=\sinh\mathcal{U}
/(\cosh\mathcal{U}-\cos\mathcal{T})$, then $\p_rF=(1-\cos\mathcal{T}\cosh\mathcal{U})
/(\cosh\mathcal{U}-\cos\mathcal{T})^2/2\cdot\p_rX/X$, where $X=\xi_+/\xi_-$. This gets rid of
the square roots and powers of $1/2(1+im/k)$ that would make the calculation much harder. Note also
that since $(\nabla^2+2k^2)\Phi=0$, we need to consider only $\nabla^2(\Phi\cdot F)$ with at least
one derivative hitting $F$.}. It is a real
function,  symmetric in $x_r,x_s$. One can check that it has a pole as $x_s\ra x_r$, as it
should, and that it doesn't have a pole as $\vec{x}_s\ra\vec{x}_r$ if $\psi_r\neq\psi_s$.
Taking the limit of zero cosmological
constant ($k\ra 0$) we see that $\Delta\ra 4\mathrm{w}^2$ (from (\ref{w})), $\gamma\ra 1$ and 
\begin{equation}
\xi_{\pm}\longrightarrow (r^2+s^2\pm 2\mathrm{w})e^{\pm m^2\mathrm{w}}\;.
\end{equation}
A brief calculation shows that this  indeed gives the Green's function for 
euclidean Taub--NUT (\ref{GreensTN}).

All previous calculations were for the oblate Pedersen metric
 (\ref{Pedersenoriginal}). One could perform an analogous calculation for the prolate
case, but it is far simpler to analytically continue the parameter
$m$ as in (\ref{continuation}) to obtain the Green's function:
\begin{equation}
G_1^{(\mu,k)}=C\cdot\frac{(1-k^2r^2)(1-k^2s^2)}{\sqrt{D}}\frac{q_+-q_-}
{q_++q_--2rsc \vone}\;\;,
\end{equation}
where now
\begin{equation}
q_\pm=(r^2+s^2\pm\sqrt{D}+2\mu kr^2s^2\vtwo)(1\pm\mu k\sqrt{D}-\mu^2k^2r^2s^2(2\vtwo-1))
^{\half\left(1-\frac{\mu}{k}\right)}
\end{equation}
(here $D=(r^2-s^2)^2-4r^2s^2(\vtwo-1)+4\mu^2k^2r^4s^4\vtwo(\vtwo-1)$) and 
\begin{equation}
c=\left[(1-\mu^2k^2r^2)(1-\mu^2k^2s^2)\right]^{\frac{1}{4}\left(1-\frac{\mu}{k}\right)}
\sqrt{(1+\mu k r^2)(1+\mu k s^2)}\;\;.
\end{equation}
It is easy to verify that $G_1^{(\mu,k)}$ reduces to the $G_1^{(0,k)}$ and $G_1^{(k,k)}$ of 
section 4 when taking the appropriate limits ($\mu\ra 0$ and $\mu\ra k$ respectively). 

Finally we can use the standard identification between the bulk--to--bulk and 
bulk--to--boundary Green's functions to obtain (in the prolate case)\footnote{As in section 
4 we normalise the bulk--to--boundary propagator to $1$ when $r=0$.}
\begin{equation}
K_1^{(\mu,k)}=\lim_{s\ra1/k}\left\{\frac{G_1^{(\mu,k)}}{(1-k^2s^2)}\right\}
=\frac{(1-k^2r^2)}{\sqrt{{D^0}}}\frac{q^0_+-q^0_-}
{q^0_++q^0_--2krc^0\vone}
\end{equation}
where now
\begin{equation}
q^0_{\pm}=(1+k^2r^2\pm\sqrt{{D^0}}+2\mu k r^2\vtwo)
\left(1\pm\frac{\mu}{k}\sqrt{{D^0}}-\mu^2r^2(2\vtwo-1)\right)^{\half\left(1-\frac{\mu}{k}\right)}
\;,
\end{equation}
with ${D^0}=(1-k^2r^2)^2-4k^2r^2(\vtwo-1)+4\mu^2k^2r^4\vtwo(\vtwo-1)$ and
\begin{equation}
c^0=\left[(1-\mu^2k^2r^4)\left(1-\mu^2/k^2\right)\right]^{\frac{1}{4}\left(1-\frac{\mu}{k}\right)}
\sqrt{\left(1+\mu/k\right)(1+\mu k r^2)}\;.
\end{equation}
We can check that as $r$ approaches the boundary, 
$K_1^{(\mu,k)}\ra(1-k^2r^2)^2\delta^3(\theta,\varphi,\psi)$. So $K_1^{(\mu,k)}$ indeed
generalises the $\HH$ bulk--to--boundary propagator to QTN. 

 Having obtained this result, we could now proceed to define boundary correlation functions
by emulating Witten's arguments in \cite{Witten98} or, if we wanted to be careful about their
normalisation, by applying the methods of holographic renormalisation. To do that,
 we would need to convert our results to the Fefferman--Graham coordinate system 
\cite{FeffermanGraham85}, regulate the action and add boundary counterterms to cancel 
divergences, as
explained (for scalars in a fixed gravitational background) in \cite{deHaroetal01}. (See also
\cite{Skenderis02}, Section 5.7 for an illustrative example.) This is currently under
investigation (\cite{Zoubosinprep}).

\section{Comments on the boundary CFT}

In this section we indicate what one could learn about the boundary CFT from the
construction of the conformal QTN Green's function. We don't provide any new
results, just motivation
 for what should be a much more extensive treatment \cite{Zoubosinprep}.

Little is known about conformal field theories in $d=3$ \footnote{Of course, in flat space, 
conformal invariance allows us to go quite far. See e.g. 
\cite{Dobrevetal77,OsbornPetkos94,FradkinPalchik96,FradkinPalchik98}.}. In the best--known case
(AdS$_4\times \Srm^7$) where $\Ncal=8$ supersymmetry is preserved, it is
believed that the dual CFT is the infrared limit of large $N$, $\Ncal=8$ SYM theory. 
It is natural to conjecture that the strongly coupled CFT dual to QTN is simply the
infrared limit of large $N$, $d=3$ Yang--Mills theory (or, more precisely, the field theory
living on the worldvolume of $N$ M2--branes), but now defined on a squashed
$\Srm^3$ rather that the round one. Note that the conformal group of the squashed sphere
(i.e. the maximal group preserving the lightcones) seems to be in fact equal to the isometry
group $\SU(2)\times\Urm(1)$.  

Another, perhaps more useful, way of looking at the dual
theory is as a (non--supersymmetric) deformation of the $\Ncal=8$ theory\footnote{This 
interpretation was suggested to me by K. Skenderis.}: Since 
the metric deformation from $\AdS$ to QTN is a classical background satisfying the boundary
conditions, we can interpret it as a source coupling to the dual CFT operator
(as is standard in AdS/CFT, e.g. \cite{Balasubramanianetal98}), which for
the transverse--traceless mode is the CFT stress--tensor\footnote{The 
deformation $h_{\mn}=g_{\mn}^{(\mu,k)}-g_{\mn}^{(0,k)}$, linearised in $\mu$,
is not tranverse and traceless, having also longitudinal and trace parts,
whose dual operators also have to be added to the CFT lagrangian. It is massless in the
$\AdS$ background, 
$\Delta_L h_\mn+2\nabla_{(\mu}\nabla^\rho h_{\nu)\rho}-\nabla_{(\mu}\nabla_{\nu)}h^\rho_{\;\rho}
+6k^2h_\mn=0$, so the dual operators will have marginal conformal dimension $\Delta=3$. These
terms will, however, break the $\SO(1,4)$ conformal group to its $\SU(2)\times \Urm(1)$ subgroup.}. 
In this way we could indeed interpret this deformation as a kind of RG--flow, thinking  of the field
theory as still living on the round sphere (and thus, essentially, on flat space) 
but with extra terms in the Lagrangian. 
For the moment, however,  we will continue to consider the boundary as a squashed $\Srm^3$.

Notice that if one agrees to keep 
the $\Srm^7$ part of the compactification 
unchanged, the masses of the bulk fields should also not change
as one passes from AdS to QTN (since the cosmological constant is the same). However, given
that the boundary first becomes negatively curved and then 
degenerates as we take the limit where QTN gives the Bergman space, it 
is not clear whether we can simply use the existing
AdS/CFT dictionary when relating bulk masses to conformal dimensions of dual
operators. Assuming that there is a well--defined correspondence, we examine a few features
that we expect to be generic.

The first question we should answer is whether the bulk--to--bulk and bulk--to--boundary
propagators we have computed are of any use in ``realistic'' situations. To rephrase
the question, are there any conformally coupled scalars in the Kaluza--Klein reduction
of eleven--dimensional supergravity down to $\AdS_4$ (and thus presumably also QTN)? 
Fortunately, as is well known \cite{Biranetal84,Castellanietal84}, the answer is positive.
(Actually, there are \emph{three} conformally coupled fields, two scalars and a pseudoscalar.)

The usual AdS/CFT dictionary associates to a bulk scalar field of mass $M$ 
the conformal dimension of its dual operator through the formula
\begin{equation} \label{Delta}
\Delta_\pm=\frac{d}{2}\pm\sqrt{\frac{d^2}{4}+\frac{M^2}{k^2}}\;.
\end{equation}

What is the dual operator to our scalar field (of  $M^2=-2k^2$) in
the (known) case of AdS$_4$?
To answer this question we make use of the results of \cite{KlebanovWitten99b}. 
They considered the implications for AdS/CFT of the well--known fact 
\cite{BreitenlohnerFreedman82a,BreitenlohnerFreedman82b} that for fields of mass
\begin{equation}
-\frac{d^2}{4}<\frac{M^2}{k^2}<-\frac{d^2}{4}+1
\end{equation}
in AdS space, there are two possible quantisations.
It is clear that our case ($d=3,M^2=-2k^2$) falls in this category. 
As mentioned in Section 4.1, the two possibilities are linked to the existence 
(in this mass range) of two AdS--invariant solutions ($G_1^{(0,k)},G_2^{(0,k)}$)
to the laplace equation for $\HH$ (\ref{hyperAdS}), depending on the
boundary conditions we wish to impose.  From the boundary CFT perspective, this corresponds to
the fact that unitarity allows not only operators of conformal dimension
$\Delta_+$, as is usual in the AdS/CFT correspondence, but also $\Delta_-$. 
 We will briefly review how the correspondence works in this case (see \cite{Balasubramanianetal99,KlebanovWitten99b,MuckViswanathan99} for more details, and also \cite{Dobrev99} for a more 
group--theoretical viewpoint).

In the usual ($\Delta_+$) correspondence, a bulk field $\Phi$ is related to its boundary
value $\Phi_0$  through the bulk--to--boundary propagator $K_{\Delta_+}$ by
\cite{Witten98}:
\begin{equation} \label{btbreg}
\Phi(r,\theta,\varphi,\psi)=C\cdot \int\diff \Omega' K_{\Delta_+}^{(0,k)}(r,\theta,\varphi,\psi,\theta',\varphi',\psi')
\Phi_0^r(\theta',\varphi',\psi')\;.
\end{equation}
 This leads to the regular boundary behaviour $\Phi\ra (1-k^2r^2)^{d-\Delta_+}\Phi_0^r$ and, via
the standard AdS/CFT formula\footnote{We use the notation $\Omega$ to abbreviate the
boundary coordinates $\theta,\varphi,\psi$.}
\begin{equation} \label{AdSCFT}
e^{-I[\Phi(r,\Omega)]}
=\langle e^{\int\diff\Omega' \mathcal{O}_{\Delta_+}(\Omega')\Phi_0^r(\Omega')}\rangle
\end{equation}
(where $\Phi$ satisfies its boundary condition) 
to correlation functions for the dual boundary operator $\mathcal{O}_{\Delta_+}$ 
of conformal dimension $\Delta_+$\footnote{One should keep in mind that (\ref{AdSCFT}) relates
infinite quantities, and to get correct results one needs to regularise and renormalise by 
adding counterterms as e.g. in \cite{deHaroetal01}.}. As suggested in \cite{KlebanovWitten99b} 
and further
explained in \cite{MuckViswanathan99}, to treat the irregular boundary condition 
$\Phi\ra (1-k^2r^2)^{d-\Delta_-}\Phi_0^i$ (which is achieved through a different 
bulk--to--boundary propagator $K_{\Delta_-}$, using
\begin{equation} \label{btbirreg}
\Phi(r,\theta,\varphi,\psi)=C\cdot \int\diff \Omega' K_{\Delta_-}^{(0,k)}(r,\theta,\varphi,\psi,\theta',\varphi',\psi')
\Phi_0^i(\theta',\varphi',\psi')\;.)
\end{equation}
one needs to use a different AdS/CFT formula:
\begin{equation}
e^{-J[\Phi(r,\Omega)]} \label{AdSCFTirreg}
=\langle e^{\int\diff\Omega' \mathcal{O}_{\Delta_-}(\Omega')\Phi_0^i(\Omega')}\rangle\;.
\end{equation}
Here the functional $J$ is related to the usual action functional $I$ by a Legendre 
transformation. This is then the generator of correlation functions
 for $\mathcal{O}_{\Delta_-}$.
Furthermore, in \cite{MuckViswanathan99}, M\"uck and Viswanathan showed that to
calculate higher--point functions of $\mathcal{O}_{\Delta_-}$, one needs to define a
different bulk--to--bulk propagator $G_{\Delta_-}$ for $\Phi$, related to the usual one 
($G_{\Delta_+}$) in a 
canonical way. This is the propagator that will give $K_{\Delta_-}$ in the appropriate
limit, $G_{\Delta_-}\longrightarrow (1-k^2r^2)^{\Delta_-}K_{\Delta_-}$ as $r\ra1/k$.

In our case conformal coupling leads to $\Delta_+=2$ and $\Delta_-=1$. 
Looking back at (\ref{KHone}) and (\ref{KHtwo}) it is clear that using $K_1^{(0,k)}$ in 
(\ref{btbirreg}) leads to 
$\Phi\ra (1-k^2r^2)^2\Phi_0^i$ as $r$ approaches the boundary, while using $K_2^{(0,k)}$ in
(\ref{btbreg})
leads to $\Phi\ra (1-k^2r^2)^1\Phi_0^r$. Thus we recognise the source $\Phi_0^i$ associated to 
$K_1^{(0,k)}$ as dual to a boundary operator of conformal dimension $\Delta_-=1$, and 
similarly the source $\Phi_0^r$ linked to $K_2^{(0,k)}$ corresponds to a boundary 
operator of dimension $\Delta_+=2$ (these will be relevant operators in the CFT). 
Of course for minimal coupling ($M^2=0$) we simply
use the standard prescription (\ref{btbreg}), (\ref{AdSCFT}) to define a marginal 
boundary operator $\mathcal{O}_3$ with conformal dimension $\Delta_+=3$.

What are these boundary operators? 
The free field content of the dual $N=8$ SYM theory is (assuming one dualises the 
Yang--Mills field to a scalar, which can be done in the free theory) 
8 scalars $X^i\;,\;i=1,\ldots,8$ and eight fermions $\grl_a\;,\;a=1,\ldots,8$ (we follow
the notation of \cite{DHokerPioline00}).  The $X^i$, $\grl_a$ have canonical dimension $1/2$, $1$ 
respectively. 
In the strong coupling limit where AdS/CFT is expected to apply, we can only really
know the spectrum of the chiral operators, which are BPS and thus their dimensions are
protected. So we can follow those operators back to weak coupling and express
them in terms of the free fields.

The matching of Kaluza--Klein fields to CFT operators can be found in
\cite{Halyo98,Aharonyetal98,Minwalla98b,EntinGomis98,DHokerPioline00}. 
We find that the operator with conformal
dimension $\Delta=1$, corresponding to a conformally coupled bulk scalar, 
is $\mathcal{O}_1=\mathrm{Str}X^{i_1}X^{i_2}$ in terms of free fields\footnote{The notation 
$\mathrm{Str}$ denotes the symmetrised trace over the colour indices.}. There are two 
operators of conformal dimension $\Delta=2$, given by 
$\mathcal{O}_2=\mathrm{Str}X^{i_1}\cdots X^{i_4}$ and
$\mathcal{O}_2'=\mathrm{Str}\grl_{a_1}\grl_{a_2}$. The operator $\mathcal{O}_2$ corresponds
to a bulk scalar, while $\mathcal{O}_2'$ corresponds to a bulk pseudoscalar.
As for the minimally coupled bulk scalar, it couples to 
$\mathcal{O}_3=\mathrm{Str}X^{i_1}\cdots X^{i_6}$, and a minimally coupled pseudoscalar
to $\mathcal{O}_3'=\mathrm{Str}\grl\grl XX$.
 
Thus we have seen how one matches some operators of low dimension in the $3d$ CFT to
the corresponding bulk--to--boundary propagators $K_1^{(0,k)},K_2^{(0,k)},K_3^{(0,k)}$. 
Now we want
to briefly leave $\HH$ and look at the more general $K_1^{(\mu,k)}$ that we constructed
in section 7. It was obtained as the limit of $(1-k^2r^2)^{-1}G_1^{(\mu,k)}$, where $G_1^{(\mu,k)}$ is the
bulk--to--bulk propagator corresponding to conformal coupling. Substituting $K_1^{(\mu,k)}$ in 
(\ref{btbirreg}) we obtain a scalar mode satisfying the irregular boundary condition. 
So it is tempting to identify the dual operator to a conformally coupled scalar
on QTN, satisfying the $\Delta_-$ boundary condition,
with the same $\mathcal{O}_1$ above, via this bulk--to--boundary 
propagator. However we now lack the protection of supersymmetry, so there is
no guarantee that the dimension of $\mathcal{O}_1$ will be immune to quantum 
corrections. All we can say is that there should exist, in the dual $3d$ QFT on the
squashed three--sphere, an operator of conformal dimension $\Delta=1$, that reduces
to $\mathcal{O}_1$ in the $\mu=0$ limit. 

Since we have not found expressions for  bulk--to--bulk and bulk--to--boundary 
propagators for $\Delta>1$, we cannot say much about those cases. However, we can make
a comment just by looking at the other relatively well--understood endpoint, corresponding to 
$\mu=k$. 
We observe that although we can follow the $K_1^{(\mu,k)}$ 
bulk--to--boundary propagator (and thus the $\Delta=1$ operator)
all the way from $\HH$ to $\Bergman$, the same does not seem to be true  
for the $K_2^{(\mu,k)}$ propagator. Indeed, in the Bergman limit $K_2^{(k,k)}$ comes not
from the conformally coupled scalar but from the minimal one. 
So it seems that an operator of $\Delta=3$ on the boundary of 
$\HH$ (i.e. minimal coupling in the bulk) will end up as an operator
of $\Delta=2$ on the boundary of $\Bergman$ \footnote{In stating this we assume that the 
standard relation between the fall--off at the boundary and the dual conformal dimension
continues to hold. This may not be the case, however, since the authors of 
\cite{Britto--Pacumioetal99}, using a
group--theoretical analysis, do associate the massless bulk mode in the Bergman space 
to a $\Delta=3$ boundary operator.}, while 
the $\Delta=2$ operator at $\mu=0$
somehow vanishes from the spectrum at $\mu=k$ (or somewhere along the way)\footnote{Perhaps this
is related to the breaking of supersymmetry: The fermions in, say, $\mathcal{O}_2'$ might now
get a mass so we don't see them in the infrared.}. The fate of this mode is probably related to the
interpretation of the second solution we found for conformal coupling in section \ref{BergGreens}.   
A better understanding of this issue will require finding the QTN Green's functions
for arbitrary coupling \cite{Zoubosinprep}, and checking their behaviour as the degenerate
$\mu=k$ limit is reached. Perhaps the relation \cite{MuckViswanathan99}
between $G_1^{(0,k)}$ and $G_2^{(0,k)}$ holds also in the QTN case, thus providing an indirect
way of finding $G_2^{(\mu,k)}$.

\section{Conclusions}

In this article we made a first step towards understanding the $d=3$ conformal
field theory dual to quaternionic Taub--NUT, by calculating the bulk--to--bulk
and bulk--to--boundary Green's functions for the simplest case, that of 
a conformally coupled scalar. 
The next step in our construction would be to actually compute 
correlation functions for $\mathcal{O}_1$ on the Berger sphere 
using (\ref{AdSCFTirreg}) supplemented by the methods of holographic renormalisation. 
We leave this crucial step for the future.
(Of course, to do conformal field theory on a $d$--dimensional manifold $X_d$ equipped with
a given conformal structure, one must sum over the contributions of all
$d+1$--dimensional manifolds with negative curvature that induce this 
conformal structure on their boundary. In the case of the squashed sphere, there
are parameter ranges where, apart from QTN we also have AdS--Taub--Bolt as the ``filling
in'' manifold \cite{Hawkingetal99,Chamblinetal99}. So a full understanding of large--$N$ 
CFT on the squashed 3-sphere would require
analogous results for AdS--Taub--Bolt --probably a much more difficult problem since
it is not self--dual.)

It will be important to extend our results to the minimally
 coupled case. One could also envisage generalising to scalars of arbitrary mass, and
also to fields of higher spin.
QTN admits fermions, but as mentioned we expect the $\mathcal{N}=8$ supersymmetry
of the $\mu=0$ case to be broken for $\mu\ne0$\footnote{However, solutions with  
NUT charge are not entirely hostile to supersymmetry. See \cite{Alonso--Albercaetal00}
for a discussion of supersymmetric Kerr--Newman--AdS-Taub--NUT solutions in the
context of $\SO(2)$ gauged $d=4$ supergravity.}.

Note that there exist some results for conformally coupled 
scalar (and spinor) effective actions on the squashed
three--sphere \cite{Dowker99} and on lens spaces \cite{DeFranciaetal01}, especially for the 
extreme oblate case. However there does not yet seem to be a direct connection between these 
weak--coupling results and what could possibly be investigated using AdS/CFT.

QTN is just a special case of a whole class of quaternionic K\"ahler manifolds
with negative cosmological constant constructed by Galicki \cite{Galicki91}.
Just as QTN reduces to ETN in the Ricci--flat limit, there are
quaternionic K\"ahler manifolds that reduce to the multicentre Taub--NUT spaces (see also
\cite{Casteilletal02} for an explicit quaternionic extension of double Taub--NUT). The
boundary of those spaces is again a squashed three--sphere,
but this time with identifications, i.e. a lens space of the form $S^3/\Zset_k$.
There also exist higher dimensional versions of AdS--Taub--NUT which
could possibly be studied in a similar way \cite{AwadChamblin00}.
 
 It would be very interesting to find a string or  M--theory system that 
reproduces our model in the supergravity limit, by which we mean a configuration of
branes (and, probably, Kaluza--Klein monopoles)
that gives QTN as part of the near horizon limit. AdS$_4$ can, as we know, be 
reproduced by a stack of M2 branes in the near horizon limit. The reason the KK
monopole solution \cite{Sorkin83,GrossPerry83} might be a useful starting point is of course
that it is related to ETN, i.e. the Ricci--flat limit of QTN. In the end, one would really like
an extension of the techniques of \cite{Gauntlettetal97} (for instance) to the quaternionic
K\"ahler case.
 As for the Bergman space, it has made an appearance in the context of untwisting 
the AdS$_5\times\Srm^5$ solution of IIB supergravity into a 
$\Srm^1\times \widetilde{\Cset\Prm^2}\times \Srm^5$ solution in type IIA$^*$
supergravity, by performing a T--duality in the time direction 
\cite{Popeetal00,KehagiasRusso00}\footnote{It is worth noting that this is a BPS solution
(being related to one by T--duality), although at first sight supersymmetry seems to be lost.
As explained in \cite{Popeetal00}, the full string theory knows about supersymmetry, 
but the supergravity limit doesn't exhibit it.}. 

 It is clear that many things remain to be done to fully understand holography on QTN.
The ultimate goal would be to answer some of the questions posed in the introduction, while
in the process learning more about holography through a concrete, non--trivial realisation.

\section*{Acknowledgments}
I would like to acknowledge useful discussions with R. Roiban and D. Vaman. I would especially
like to thank K. Skenderis for communications and very helpful comments on the manuscript,
and also  M. Ro\v cek for guidance and constant 
encouragement. I am grateful to the State Scholarships Foundation of Greece for financial
support during a large part of this project.

\appendix

\section{Pedersen vs. AdS--Taub--NUT}

In this article we make use of the ``polar--like'' Pedersen metric for  
quaternionic Taub--NUT. This is because in these coordinates both
the AdS and Bergman limits are easy to work with. To gain contact
with recent literature on AdS--Taub--NUT \cite{Hawkingetal99,Chamblinetal99,Emparanetal99,BehrndtDallAgata02}, we show how to relate the metric used in those references to the Pedersen metric. 

Substituting (\ref{AdSTNcond}) in (\ref{AdSTN}) we obtain the 
most commonly used metric for AdS--Taub--NUT:
\begin{equation} \label{AdSTNapp}
\diff s^2=\frac{\diff r^2}{V(r)}+V(r)[\diff\tau+2n\cos{\theta}\diff\varphi]^2
+(r^2-n^2)[\diff\theta^2+\sin^2{\theta}\diff\varphi^2]\;,
\end{equation}
where 
\begin{equation}
V(r)=\frac{r-n}{r+n}\left[1+k^2(r-n)(r+3n)\right]\;,
\end{equation}
and $\tau=2n\psi$, $k^2=-\Lambda/3$. In this form, there is a nut at
$r=n$.

To see that this is the same as the Pedersen metric, we can make the change
\begin{equation}
r\rightarrow r+n
\end{equation}
(transporting the nut to $r=0$), followed by the transformation\footnote{This is a slight
adaptation of eq. (2.8) in \cite{Chamblinetal99}.}
\begin{equation}
r=\frac{\rho^2/(2n)}{1-k^2\rho^2}\;.
\end{equation}
Then, identifying $\mu^2/k^2=1-1/(4k^2n^2)$ (and relabelling $\rho\ra r$) 
brings the metric to the prolate form (\ref{Pedersen}).
We see that the Ricci--flat boundary case ($\mu^2=3/4k^2$) corresponds to a nut
charge $n=1/k$, while to reach the Bergman limit ($\mu^2=k^2$) we need to 
take $n\ra \infty$. At $n=1/(2k)$ we recover $\AdS$, and for $n<1/(2k)$ we find ourselves in the
oblate case.

Taking $k=0$ brings us to the relation between the more commonly used metric 
for euclidean Taub--NUT ((\ref{AdSTNapp}) with $V(r)=(r-n)/(r+n))$ to the one we use in
section 3.1, and relates the nut charges of these two metrics.

\section{Coordinate systems on hyperbolic space} \label{Coords}

 In the AdS/CFT correspondence, it is usually convenient to write the metric
for euclidean AdS in Poincar\'e (upper half plane) coordinates. Then the conformal
flatness of the boundary metric is obvious. In our case, since we are interested
in deforming away from the conformally flat case, it is more useful to exhibit the
$\SU(2)\times\SU(2)$ symmetry of the boundary by using polar--type coordinates for 
hyperbolic space. Then it is clear how this bi-invariant case arises as a special
case of the squashed $\SU(2)\times\Urm(1)$ metric. 
Since the bulk of the AdS/CFT literature on propagators uses the Poincar\'e 
parametrisation, we give a short explanation of our coordinates and the form 
of the propagators. 

 We consider $\HH$ as the hyperboloid embedded in $\Rset^{1,4}$: 
\begin{equation} \label{Hyperboloid}
X_0^2-X_1^2-X_2^2-X_3^2-X_4^2=\frac{1}{k^2}\;\;,
\end{equation}
where $1/k^2$ is the radius of curvature. To obtain the Poincar\'e 
parametrisation we solve (\ref{Hyperboloid}) by:
\begin{equation}
\begin{split}
X_0&=\frac{y}{2}\left(1+\frac{1}{y^2}\left(x^2+\frac{1}{k^2}\right)\right)\;,\\
X_a&=\frac{x_a}{ky}\;\;,\quad\quad\quad\;\;\; a=1,2,3\;\;,\\
X_4&=\frac{y}{2}\left(1+\frac{1}{y^2}\left(x^2-\frac{1}{k^2}\right)\right)\;,
\end{split}
\end{equation}
where $x^2=\sum_{a=1}^3x_a^2$ (see e.g. \cite{Aharonyetal00}). This gives the metric 
\begin{equation}
\diff s^2=-\diff X_0^2+\sum_{i=1}^{4}\diff X_i^2=\frac{1}{k^2y^2}
\left(\diff y^2+\diff x_1^2+\diff x_2^2+\diff x_3^2\right)\;\;.
\end{equation}
In this parametrisation the boundary is at $y=0$. 

On the other hand, to obtain the ``polar/stereographic'' parametrisation of
(\ref{HH1}) we solve (\ref{Hyperboloid}) as follows (e.g. \cite{Petersen99}):
\begin{equation} \label{polar}
\begin{split}
X_0&=\frac{1}{k}\frac{1+k^2w^2}{1-k^2w^2}\;\;,\\
X_i&=\frac{2w_i}{1-k^2w^2}\;\quad,\;\; i=1,2,3,4\;\;,
\end{split}
\end{equation}
where here $w^2=\sum_{i=1}^{4}w_i^2$. Then we find
\begin{equation}
\diff s^2=-\diff X_0^2+\sum_{i=1}^{4}\diff X_i^2=
\frac{4}{1-k^2w^2}\left(\diff w_1^2+\diff w_2^2+\diff w_3^2+\diff w_4^2\right)\;.
\end{equation}
First we convert to the complex coordinates $z_1=w_1+iw_2,z_2=w_3+iw_4$ to obtain
\begin{equation} \label{complex}
\diff s^2=\frac{4}{1-k^2(z_1\bar{z}_1+z_2\bar{z}_2)}
\left(\diff z_1\diff \bar{z}_1+\diff z_2\diff\bar{z}_2\right)\;\;,
\end{equation}
and finally we convert to polar coordinates using 
\begin{equation} \label{polarfinal}
z_1=r \cos{\frac{\tr}{2}}e^{\frac{i}{2}(\psi_r+\fr)}\;\;,\;\;
z_2=r \sin{\frac{\tr}{2}}e^{\frac{i}{2}(\psi_r-\fr)}\;
\end{equation}
to obtain the metric in (\ref{HH1}):
\begin{equation}
\diff s^2=\frac{4}{(1-k^2r^2)^2}\left[\diff r^2+r^2(\sone^2+\stwo^2+\sthree^2)\right]\;.
\end{equation}
Now the boundary is at $r=1/k$.

We are interested in the expression for the chordal distance in the two 
coordinate systems. In Poincar\'e coordinates we have the expression
(e.g. \cite{DHokeretal99b}): 
\begin{equation}
u=\frac{(y-y')^2-(\vec{x}-\vec{x}')^2}{2yy'}\;\;,\;\; 
(\vec{x}-\vec{x}')^2=\sum_{a=1}^{3}(x_a-x_a')^2\;.
\end{equation}
First we convert this to the homogeneous $X_0,...,X_4$ coordinates
using
\begin{equation}
y=\frac{1}{k^2(X_0-X_4)}\;\;,\;\;x_a=kyX_a,\;\;a=1,2,3\;
\end{equation}
to obtain, as expected:
\begin{equation}
u=\left(k^2(X_0X_0'-\sum_{i=1}^{4}X_iX_i')-1\right)\;.
\end{equation}
Now we convert to the $w_i$  coordinates using (\ref{polar}):
\begin{equation}
u=\frac{2k^2\left(w^2+w^{\prime2}-2\sum_{i=1}^{4}w_iw_i'\right)}
{(1-k^2w^2)(1-k^2w^{\prime2})}\;.
\end{equation}
Finally we convert to polar coordinates using (\ref{complex}) 
and (\ref{polarfinal}) (setting $z_1'=s\cos(\theta_s/2)$
$e^{\frac{i}{2}(\psi_s+\varphi_s)}$ and so forth) to obtain the chordal
distance for hyperbolic space:
\begin{equation}
u^{(\mu=0,k)}=2k^2\frac{r^2+s^2-rs\vone}{(1-k^2r^2)(1-k^2s^2)}\;.
\end{equation}
In this way we have recovered the expression in Section 4.1.

\bibliography{References}
\bibliographystyle{JHEP}

\end{document}